\documentstyle[12pt,epsf,epsfig,fleqn]{article}
\newcount\timecount
\newcount\hours \newcount\minutes  \newcount\temp \newcount\pmhours

\hours = \time
\divide\hours by 60
\temp = \hours
\multiply\temp by 60
\minutes = \time
\advance\minutes by -\temp
\def\hour{\the\hours}
\def\minute{\ifnum\minutes<10 0\the\minutes
            \else\the\minutes\fi}
\def\clock{
\ifnum\hours=0 12:\minute\ AM
\else\ifnum\hours<12 \hour:\minute\ AM
      \else\ifnum\hours=12 12:\minute\ PM
            \else\ifnum\hours>12
                 \pmhours=\hours
                 \advance\pmhours by -12
                 \the\pmhours:\minute\ PM
                 \fi
            \fi
      \fi
\fi
}

\def\monthname{\relax\ifcase\month 0/\or January\or February\or
   March\or April\or May\or June\or July\or August\or September\or
   October\or November\or December\else\number\month/\fi}

\def\bold#1{\setbox0=\hbox{$#1$}%
     \kern-.025em\copy0\kern-\wd0
     \kern.05em\copy0\kern-\wd0
     \kern-.025em\raise.0433em\box0 }

\def\ga{\mathrel{\raise.3ex\hbox{$>$\kern-.75em\lower1ex\hbox{$\sim$}}}}
\def\la{\mathrel{\raise.3ex\hbox{$<$\kern-.75em\lower1ex\hbox{$\sim$}}}}
\def\gev{{\rm \, Ge\kern-0.125em V}}
\def\tev{{\rm \, Te\kern-0.125em V}}
\def\beq{\begin{equation}\!\!\!\!\!\!\!\!\!\!\! }
\def\eeq{\end{equation}}
\def\bdm{\begin{displaymath}\!\!\!\!\!\!\!\!\!\!\! }   
\def\edm{\end{displaymath}}
\def\beqar{\begin{eqnarray}}
\def\eeqar{\end{eqnarray}}

\def\m12{m_{1\!/2}}

\def\ga{\mathrel{\raise.3ex\hbox{$>$\kern-.75em\lower1ex\hbox{$\sim$}}}}
\def\la{\mathrel{\raise.3ex\hbox{$<$\kern-.75em\lower1ex\hbox{$\sim$}}}}
\def\gyr{{\rm \, G\kern-0.125em yr}}
\def\gev{{\rm \, Ge\kern-0.125em V}}
\def\tev{{\rm \, Te\kern-0.125em V}}

\def\m12{m_{1\!/2}}

\newcommand\iso[2]{\mbox{${}^{#2}${\rm #1}}}
\def\he#1{\iso{He}{#1}}
\def\be#1{\iso{Be}{#1}}
\def\li#1{\iso{Li}{#1}}
\def\b1#1{\iso{B}{1#1}}

\newcommand\pfrac[2]{\left( \frac{#1}{#2} \right)}
\newcommand\pref[1]{(\ref{#1})}

\newcommand\EE[2]{{#1}\!\times\! 10^{#2}}

\def\gappeq{\mathrel{\rlap {\raise.5ex\hbox{$>$}}
{\lower.5ex\hbox{$\sim$}}}}

\def\lappeq{\mathrel{\rlap{\raise.5ex\hbox{$<$}}
{\lower.5ex\hbox{$\sim$}}}}

\def\Toprel#1\over#2{\mathrel{\mathop{#2}\limits^{#1}}}

\begin{document}

\begin{titlepage}
\pagestyle{empty}
\baselineskip=21pt
\rightline{astro-ph/0211258}
\rightline{CERN--TH/2002-207}
\rightline{UMN--TH--2115/02}
\rightline{TPI--MINN--02/44}
\vskip .5in
\begin{center}

{\large {\bf
Updated Nucleosynthesis Constraints on Unstable Relic 
Particles}}

\end{center}   
\begin{center}   
\vskip 0.2in   
{{\bf Richard H.~Cyburt}$^1$, {\bf John Ellis}$^2$, {\bf Brian
D.~Fields}$^3$ and {\bf  Keith A.~Olive}$^{4}$}\\
\vskip 0.1in
{\it
$^1${Department of Physics, University of Illinois,
Urbana, IL 61801}\\
$^2${Theory Division, CERN, CH 1211 Geneva 23, Switzerland}\\
$^3${Department of Astronomy, University of Illinois,
Urbana, IL 61801}\\
$^4${Theoretical Physics Institute,
University of Minnesota, Minneapolis, MN 55455, USA}}\\
\vskip 0.2in
{\bf Abstract}
\end{center}
\baselineskip=18pt \noindent


We revisit the upper limits on the abundance of unstable massive relic
particles provided by the success of Big-Bang Nucleosynthesis
calculations. We use the cosmic microwave background data to constrain the
baryon-to-photon ratio, and incorporate an extensively updated compilation
of cross sections into a new calculation of the network of reactions
induced by electromagnetic showers that create and destroy the light
elements deuterium, \he3, \he4, \li6 and \li7.  We derive analytic
approximations that complement and check the full numerical calculations.
Considerations of the abundances of \he4 and \li6 exclude exceptional
regions of parameter space that would otherwise have been permitted by
deuterium alone. We illustrate our results by applying them to massive
gravitinos. If they weigh $\sim 100$~GeV, their primordial abundance
should have been below about $10^{-13}$ of the total entropy. This would
imply an upper limit on the reheating temperature of a few times
$10^7$~GeV, which could be a potential difficulty for some models of
inflation. We discuss possible ways of evading this
problem.

\vskip 0.5in
\vfill
\leftline{CERN--TH/2002-207}
\leftline{October 2002}
\end{titlepage}

\section{Introduction}

Some of the most stringent constraints on unstable massive particles are
provided by their effects on the abundances of the light nuclei produced
primordially in the early Universe via Big-Bang Nucleosynthesis (BBN)
\cite{Lindley:1984bg}-\cite{Kohri:2001jx}. Astrophysical determinations of
the abundances of D, \he4 and \li7 agree with those found in homogeneous
BBN calculations, for a suitable range of the baryon-to-photon ratio
$\eta$ \cite{osw,cfo1}. The decay products of massive unstable particles
such as gravitinos would have produced electromagnetic and/or hadronic
showers in the early Universe, which could have either destroyed or
created these nuclei, perturbing this concordance. Maintaining the
concordance provides important upper limits on the abundances of such
massive unstable particles. The power of this argument has recently been
increased by observations of the power spectrum of fluctuations in the
cosmic microwave background (CMB) \cite{cfo2}. These now provide an
independent determination of $\eta$ that is in rather good agreement with
the value suggested by BBN calculations \cite{dasi}-\cite{vsa}, reducing
one of the principal uncertainties in the previous BBN limits on massive
unstable particles.

This development has triggered us to re-evaluate these BBN constraints. We
do so via a new calculation of the network of nuclear reactions induced by
electromagnetic showers that create and destroy the light elements
deuterium (D), \he3, \he4, \li6 and \li7. We perform a coupled-channel analysis
of the light-element abundances, which enables us to analyze the possible
existence of isolated `islands' of parameter space that are not minor
perturbations of standard homogeneous BBN calculations. In carrying out
this program, we also take the opportunity to improve previous
theoretical treatments of some reactions in essential ways. We also derive
analytic approximations to our results, which serve as a check of the
numerics, and offer additional insight into the essential physics.

Considerations of the D abundance alone would have led to the apparent
existence of finely-tuned `tails' of parameter space that extend the
region allowed by standard BBN calculations to rather larger abundances of
unstable relic particles. In the past, these `tails' were argued to be
excluded on the basis of the combined D + \he3 abundance. It is now widely
considered that the complicated chemical and stellar history of \he3
renders such an argument unsafe \cite{vofc}. However, we show here that
these tails are excluded robustly by the astrophysical abundance of \he4,
and also conflict with the measured \li6 abundance. These other abundances
also exclude a disconnected `channel' of parameter space that would have
been allowed by the D abundance alone. Overall, the most stringent upper
limit on the possible abundance of an unstable massive relic particle $X$
with lifetime $\tau_X \gappeq 10^2$~s is provided by the \li6 abundance.
For
$\tau_X = 10^8$~s, we find

\begin{equation}
M_X {n_X^0 \over n_\gamma^0} \; < \; \EE{5.0}{-12}~{\rm GeV},
\label{bestlimit}
\end{equation}
with the upper limit from the \he4 abundance being about two orders of 
magnitude less stringent.

As an illustrative application of this new analysis, we reconsider the
allowed abundance of unstable gravitinos ${\tilde G}$ with masses up to
about 10~TeV, which are expected to have lifetimes $\gappeq 10^2$~s. Using
standard calculations of thermal gravitino production in the early
Universe \cite{Nanopoulos:1983up} - \cite{Bolz:2000fu} ,
\cite{Kawasaki:1994af}, the constraint (\ref{bestlimit}) suggests a
stringent upper limit on the reheating temperature $T_R$ of the Universe
following inflation:
\begin{equation}
T_R \lappeq 10^6~{\rm GeV~~for}~M_{\tilde G} \sim 100~{\rm GeV}.
\label{highestreheat}
\end{equation} 
Such a stringent upper limit can be problematic for standard inflationary
models~\cite{infl}, some of which predict higher $T_R \sim 10^{12}$~GeV. A
low reheating temperature might also be problematic for some models of
baryogenesis in which thermal production of baryon-number-violating
particles is necessary, though not for others. The simplest leptogenesis
scenarios~\cite{fy} require the production of a right-handed neutrino
state whose decays violate lepton number. However, even scenarios which
require the thermal production of right-handed neutrinos can in fact
accommodate very low reheat temperatures~\cite{luty,cdo}, and non-thermal
production of particles with masses less than the inflaton mass
is possible for very low reheating temperatures~\cite{nos}. Leptogenesis
scenarios~\cite{Asaka:1999yd} of this type as well as those involving
preheating~\cite{pre} have been proposed~\footnote{We also recall that
Affleck-Dine~\cite{ad} leptogenesis~\cite{cdo,adl} can be accomplished at
low $T_R$.}.

Reheat temperatures higher than (\ref{highestreheat}) would be allowed if
$\tau_{\tilde G} \lappeq 10^4$~s, as might occur if $M_{\tilde G} \ga
3$~TeV and/or the gravitino has many available decay modes.
Alternatively, one may consider the possibility of a very light gravitino
that would be stable or metastable. However, in this case the upper limit
(\ref{bestlimit}) would apply to the next-to-lightest supersymmetric
particle (NLSP). In conventional MSSM scenarios with a very light
gravitino, this might be the lightest neutralino $\chi$. However, the
relic $\chi$ abundance may also be calculated, and is likely to conflict
with the upper limit (\ref{bestlimit}).

This brief discussion serves to emphasize the importance of evaluating the
cosmological upper limit on the possible primordial abundance of unstable
relic particles, to which the bulk of this paper is devoted. In
Section~\ref{sect:rxnet}, we discuss the most relevant photodissociation
and photoproduction processes and describe their implementation in a code
to calculate the network of reactions creating and destroying light
elements in the early Universe. A brief discussion of the observational
constraints used is in Section~3. Analytical and numerical results for
light-element abundances are presented in Section~4. Our main constraints
on unstable particles are described in Section~\ref{sect:limits}, where we
compare and combine the upper limits obtained by considering different
light nuclei. Finally, in Section~\ref{sect:grav} we discuss in more
detail the implications of our results for cosmological gravitinos.  
Section~7 contains some comments on particles with longer (and shorter)  
lifetimes. In Appendix~\ref{sect:cs}, we compile and discuss the cross
sections used in this analysis.

\section{Photon Injection and Abundance Evolution}
\label{sect:rxnet}

As an example of the constraints imposed on the abundance of a heavy
metastable particle by observations of light elements, we consider the
radiative decay of a massive particle (such as a gravitino) with lifetime
$\ga 10^4$~s. The energetic decay photon initiates an electromagnetic
shower, which in turn initiates a network of nuclear interactions. The
decays of some unstable heavy particles also initiate hadronic showers,
which would provide an additional set of constraints. These are typically
important for shorter lifetimes, $\tau \lappeq 10^2$~s. However, we
restrict our attention in this paper to the network of reactions induced
by electromagnetic showers, and comment on very short (and very long)
lifetimes in Section~7.

\subsection{The Initial Degraded Photon Spectrum}

At the epoch of interest to us here, $\tau_X \gappeq 10^4$ sec, the massive
gravitino\footnote{For definiteness, we will refer to the decaying
particle as a gravitino, though our analysis is general and pertains to
any massive particle with electromagnetic decays in the lifetime range
considered.} is very non-relativistic, and can be treated as if at rest
with respect to the background. We assume that a gravitino of mass
$M_X$ decays into a photon ($\gamma$) and a neutralino ($\chi$), each
with their respective energies,
\beq
E_\gamma = \frac{M_X^2 - m_\chi^2}{2M_X} \ {\rm and} \ E_\chi = \frac{M_X^2 +
m_\chi^2}{2M_X}. 
\eeq
In the limit $M_X \gg m_\chi$, the energies become almost equal:
$E_\gamma \sim E_\chi \sim \frac{1}{2}M_X$.

The primary photon with injection energy $E_0$
interacts
with the background plasma and creates an electromagnetic cascade.
The most rapid interactions in this cascade are 
pair production
$\gamma \gamma_{\rm bg} \rightarrow e^+ e^-$ off of background
photons, and
inverse compton scattering.
These processes rapidly
redistribute the injected energy, and the nonthermal
photon spectrum rapidly reaches a
quasi-static equilibrium as discussed in~\cite{ellis,kawmoroi,psb}.  The
``zeroth generation''
quasi-equilibrium photon energy spectrum is 
\beq
\label{eq:injection}
p_\gamma(E_\gamma) = \left\{ \begin{array}{lll}
K_0\left( \frac{E_X}{E_\gamma} \right)^{1.5} &
\mbox{if $E_\gamma < E_X$} \\
K_0\left( \frac{E_X}{E_\gamma} \right)^{2.0} &
\mbox{for $E_X < E_\gamma < E_C$} \\
0 \ \ \ \ \ \ \ \ \ \ \ \ & \mbox{otherwise}
\end{array} \right\}
\eeq
where the normalization constant $K_0$ is determined by demanding that the
total energy be equal to the injected energy: $K_0 =
E_0/[E_X^2(2+\ln{(E_C/E_X)})]$.  This spectrum is the same as that used by
Protheroe, Stanev, \& Berezinksy \cite{psb} and 
Jedamzik~\cite{kartsen}, and also agrees with the result
of a detailed numerical integration of
the full Bolzmann equation by
Kawasaki and
Moroi~\cite{kawmoroi}~\footnote{The spectrum stated in~\cite{kawmoroi}
includes further photon degradation, which has been factored out to
determine our spectrum.}. It is a broken power law with a transition at
$E_\gamma = E_X$ and a high-energy cutoff at $E_\gamma = E_C$. We adopt
the same energy limits as Kawasaki and Moroi, namely $E_X = m_e^2/80T$ and
$E_C = m_e^2/22T$. Physically, these scales arise due to the competition
between photon degradation rates. The scales rise as the temperature
drops, in which case there exist more high-energy photons to break up
nuclei.

The zeroth-generation nonthermal photons then suffer additional
interactions of Compton scattering, ordinary pair production off of nuclei,
and $\gamma-\gamma$ scattering.
These slower processes further degrade the photon specturm.
The evolution of the resulting ``first-generation''
photons is
governed by 
\beq
\frac{d {\mathcal N}_\gamma}{d t}(E_\gamma ) =
\frac{n_X}{\tau_X} p_\gamma (E_\gamma) - {\mathcal N}_\gamma 
(E_\gamma )\Gamma_\gamma (E_\gamma ),
\label{eqn:ar1_1}
\eeq
where $n_X = n_X^0(1+z)^3\exp{(-t/\tau_X)}$ and $\tau_X$ are the decaying
particle number density at redshift $z$ and mean lifetime,
respectively. Also, ${\mathcal N}_\gamma$ is the photon energy spectrum,
which is simply the product of the density of states ${\mathcal D}_\gamma$
and the occupation number fraction $f_\gamma$. Integrating ${\mathcal
N}_\gamma$ over all energies yields the number density $n_\gamma^{\rm inj}$ of
the injected photons. Further, $\Gamma_\gamma$ is the rate at which the
photons are further degraded through further interactions with the
background plasma. 
The key difference between $p_\gamma$ and ${\mathcal
N}_\gamma$ is that the rates degrading photons directly after injection
are much faster than the rates that further degrade photon energy
determining ${\mathcal N}_\gamma$. We note that the effects due to the
expansion of the universe on the photon spectrum are negligible because,
during this epoch, electromagnetic interactions are much faster than the
expansion rate.

The dominant photon degradation rates are those for double photon
scattering, Compton scattering and pair production off nuclei. Because
their high rates are fast compared to the cosmic expansion, 
the photon distribution reaches quasi-static
equilibrium (QSE).  This distribution is given by setting
(\ref{eqn:ar1_1}) equal to zero, yielding

\beq
{\mathcal N}_\gamma^{\rm QSE}(E_\gamma) =
\frac{n_Xp_\gamma (E_\gamma)}{\Gamma_\gamma(E_\gamma)\tau_X}.
\eeq
This QSE solution is the same as that derived in~\cite{kawmoroi}, where
it is called $f_\gamma(\epsilon_\gamma)$. The photon spectrum $p_\gamma$
can be determined easily from this equation, knowing that double-photon
scattering dominates the high-energy region, whereas Compton scattering 
and pair production off nuclei dominate at lower energies. We recall that the
redshift dependence of this QSE solution lies entirely in $n_X$,
$p_\gamma$, and $\Gamma_\gamma$.

\subsection{Photo-Destruction and -Production of Nuclei}

The equations governing the production and destruction of nuclei are
very similar to those for photons, being given by
\beq
\frac{d {\mathcal N}_A}{d t}(E_A) = J_A(E_A) -
{\mathcal N}_A(E_A)\Gamma_A(E_A),
\label{eqn:nucrate}
\eeq
where $J_A$ and $\Gamma_A$ are the source and sink rates of primary
species $A$.  The derivative, $d/dt$ takes into account the
redshifting of energies and the dilution of particles due to the
expansion of the universe.  The source terms for the primary species
are due to the photodissociation of background particles, and are defined by: 
\beq
J_A(E_A) = \sum_T n_T\int_{0}^\infty\!\!\!\!\! dE_\gamma
{\mathcal N}_\gamma^{\rm QSE}(E_\gamma)\, 
\sigma_{\gamma + T\rightarrow A} (E_\gamma)
\,\delta \left[{\mathcal E}_A^T(E_\gamma) - E_A\right], 
\label{eqn:ar3_1} 
\eeq
where ${\mathcal E}_A^T(E_\gamma)$ is the
energy of the $A^{\rm th}$ species produced by the photodissociation
reaction $\gamma + T\rightarrow A$. The sinks are similarly defined by
\beq
\Gamma_A(E_A) = \sum_P \int_{0}^\infty\!\!\!\!\! dE_\gamma
{\mathcal N}_\gamma^{\rm QSE}(E_\gamma)\, 
\sigma_{\gamma + A\rightarrow P} (E_\gamma).
\eeq
Since we are interested in calculating total abundances of elements, it is
necessary to integrate (\ref{eqn:nucrate}) over the energy $E_A$.
The equation then becomes
\beq
\frac{dn_A}{dt} = \sum_T n_T\int_{0}^\infty\!\!\!\!\! dE_\gamma
{\mathcal N}_\gamma^{\rm QSE}(E_\gamma)\, 
\sigma_{\gamma + T\rightarrow A} (E_\gamma)
- n_A\sum_P
\int_{0}^\infty\!\!\!\!\! dE_\gamma {\mathcal N}_\gamma^{\rm QSE}(E_\gamma)\, 
\sigma_{\gamma + A\rightarrow P} (E_\gamma).
\eeq
This removes the redshifting term, leaving only the dilution term
in the derivative, $d/dt$.  It is useful to use the mole fraction $Y_i
\equiv n_i/n_{\rm B}$ of baryons in a particular
nuclide, rather than the absolute abundance.  This allows us to take
out the expansion effects, yielding
\beq
\frac{d Y_A}{d t} = \sum_T Y_T\int_{0}^\infty dE_\gamma
{\mathcal N}_\gamma^{\rm QSE}(E_\gamma)\, 
\sigma_{\gamma + T\rightarrow A} (E_\gamma)
- Y_A\sum_P
\int_{0}^\infty\!\!\!\!\! dE_\gamma {\mathcal N}_\gamma^{\rm QSE}(E_\gamma)\, 
\sigma_{\gamma + A\rightarrow P} (E_\gamma),
\eeq
where $n_B$ is the baryon number density and $d/dt$ is an ordinary
time derivative.
It is also convenient to change from time differentiation to
differentiation with respect to redshift, and also to extract the factor
$n_XE_0/\tau_X$ out of ${\mathcal N}_\gamma^{\rm QSE}$. In this way, we obtain
\beqar
\label{eq:primary}
\frac{d Y_A}{d z} & = & -\frac{rM_Xn_\gamma^0}{H_r\tau_X}\exp{\left(
\frac{-1}{H_r\tau_X(1+z)^2} \right)}  \\
 & &  \times 
\left[ \sum_T Y_T\int_{0}^\infty\!\!\!\!\!
dE_\gamma \left(\frac{\tau_X}{E_0n_X}{\mathcal
N}_\gamma^{\rm QSE}(E_\gamma)\right)\, \sigma_{\gamma + T\rightarrow A}
(E_\gamma) \right. \\
  & & \left. - Y_A\sum_P \int_{0}^\infty\!\!\!\!\! dE_\gamma
\left(\frac{\tau_X}{E_0n_X}{\mathcal
N}_\gamma^{\rm QSE}(E_\gamma)\right)\,  \sigma_{\gamma + A\rightarrow P}
(E_\gamma) \right],
\eeqar
where we have used $E_0 = \frac{1}{2}M_X$ (corresponding to two-body decay 
into a photon and a particle of negligible mass), 
and have defined $r \equiv 
n_X^0/n_\gamma^0$
and $H_r \equiv \sqrt{32\pi G\rho_{rad}^0/3}$. We should note that, with
these terms pulled out, the integrals become functions only of redshift,
and the dependence on the properties of the decaying particle has been
removed. This formulation is very useful in making the 
numerical implementation fast and
efficient.

\begin{table}
\caption{\it
The relevant photo-dissociation reactions
and their respective threshold energies are listed in the
Table below, and their cross sections are listed in 
Appendix~\ref{sect:cs}.}
\label{tab:photo-thresh}
\begin{center}
\begin{tabular}[t]{|l||r|} \hline
Reaction  & Threshold (E$_{\gamma,\rm th}$)\\ \hline\hline
d($\gamma$,n)p              &  2.2246 MeV \\ \hline\hline
t($\gamma$,n)d              &  6.2572 MeV \\
t($\gamma$,np)n             &  8.4818 MeV \\ \hline\hline
\he3($\gamma$,p)d           &  5.4935 MeV \\
\he3($\gamma$,np)p          &  7.7181 MeV \\ \hline\hline
\he4($\gamma$,p)t           & 19.8139 MeV \\ 
\he4($\gamma$,n)\he3        & 20.5776 MeV \\ 
\he4($\gamma$,d)d           & 23.8465 MeV \\
\he4($\gamma$,np)d          & 26.0711 MeV \\ \hline\hline 
\li6($\gamma$,np)\he4       &  3.6989 MeV \\
\li6($\gamma$,X)$^3$A       & 15.7947 MeV \\ \hline\hline
\li7($\gamma$,t)\he4        &  2.4670 MeV \\
\li7($\gamma$,n)\li6        &  7.2400 MeV \\
\li7($\gamma$,2np)\he4      & 10.9489 MeV \\ \hline\hline
\be7($\gamma$,\he3)\he4     &  1.5866 MeV \\
\be7($\gamma$,p)\li6        &  5.6058 MeV \\
\be7($\gamma$,2pn)\he4      &  9.3047 MeV \\ \hline
\end{tabular}
\end{center}
\end{table}

\subsection{Secondary Element Production}

In the previous Section we discussed the production of light elements by
the photo-dissociation of heavier elements.  However, the initial
photo-production/destruction of light nuclei is not necessarily the only
process that happens before thermalization. The primary interactions
produce non-thermal particles, which then interact with the background
plasma, degrading their energy. However, they may still have enough energy
to initiate further, secondary nuclear interactions.

We now modify the evolution of the primary particles described in the
previous section to include energy-degrading interactions:
\beq
\frac{d{\mathcal N}_A}{dt}(E_A) = J_A(E_A) - {\mathcal
N}_A(E_A)\Gamma_A(E_A) - \frac{\partial}{\partial E_A} \left[ b_A(E_A)
{\mathcal N}_A(E_A) \right],
\eeq
where we have added the last term to include the energy degradation of the
species $A$, where $b_A = -dE/dt$ is the rate of energy loss.
This term appears as an energy gradient, conserving particle
number in the absence of sources and sinks.

In most situations, the energy degradation rate is much faster than any
sink, so that the sinks can be ignored.  For unstable particles, if
the lifetime of the particle is comparable to the stopping time of
that species, then the $\Gamma_A$ term cannot be ignored.  In general
the interactions are fast enough to reach a quasi-static equilibrium, but
the form of the solution is somewhat more complicated than the photon 
case:
\beq
{\mathcal N}_A^{\rm QSE}(E_A) = \frac{1}{b_A(E_A)}
\int_{E_A}^\infty\!\!\!\!\! dE_A^{^\prime} \exp{ \left[ -
\int_{E_A}^{E_A^{^\prime}} dE_A^{^{\prime \prime}}
\frac{\Gamma_A(E_A^{^{\prime \prime}})}{b_A(E_A^{^{\prime \prime}})}
\right] } J_A(E_A^{^\prime}).
\eeq
Substituting $J_A$ into this equation, we get
\beq
{\mathcal N}_A^{\rm QSE}(E_A) = \frac{1}{b_A(E_A)}\sum_Tn_T \int_{{\mathcal
E}_A^{-1}(E_A)}^\infty\!\!\!\!\! dE_\gamma {\mathcal N}_\gamma(E_\gamma)
\sigma_{\gamma + T \rightarrow A}(E_\gamma)  \exp{\left[  -
\int_{E_A}^{{\mathcal E}_A(E_\gamma)} dE_A^{^{\prime \prime}}
\frac{\Gamma_A(E_A^{^{\prime \prime}})}{b_A(E_A^{^{\prime \prime}})}
\right] }.
\eeq
With this QSE solution in hand, we can determine the rate of secondary
production of an element $S$:
\beq
\left. 
\frac{d{\mathcal N}_S}{dt}\right|_{\rm sec} \!\!\!\!\!\!\! (E_S) =
\sum_{T^{^\prime}} \int_0^\infty \!\!\!\!\! dE_A
{\mathcal N}_A^{\rm QSE}(E_A) \Gamma_{A+T^{^\prime}\rightarrow S}(E_A)
\delta \left[ {\mathcal E}_S(E_A) - E_S \right].
\eeq
Integrating over $E_S$ yields:
\beq
\left. \frac{dn_S}{dt}\right|_{\rm sec} = \sum_{T^{^\prime}}
n_{T^{^\prime}}\int_0^\infty \!\!\!\!\! dE_A {\mathcal N}_A^{\rm QSE}(E_A)
\sigma_{A+T^{^\prime}\rightarrow S}(E_A) |v_A|.
\eeq
Again, using the mole baryon fraction to remove expansion effects from
the differential equation, we obtain
\beq
\left. \frac{d Y_S}{d t}\right|_{\rm sec} =
\sum_{T^{^\prime}} Y_{T^{^\prime}}\int_0^\infty \!\!\!\!\! dE_A
{\mathcal N}_A^{\rm QSE}(E_A)
\sigma_{A+T^{^\prime}\rightarrow S}(E_A) |v_A|.
\eeq

\begin{table}
\caption{\it
Relevant secondary reactions are listed below, in a format similar to 
Table~\ref{tab:photo-thresh}.
}
\label{tab:nuke-thresh}
\begin{center}
\begin{tabular}[t]{|l||r|} \hline
Reaction  & Threshold (E$_{p,\rm th}$)\\ \hline\hline
p(n,$\gamma$)d             & 0.0000 MeV \\ \hline\hline
\he4(t,n)\li6              & 8.3870 MeV \\
\he4(\he3,p)\li6           & 7.0477 MeV \\ \hline\hline
\he4(t,$\gamma$)\li7       & 0.0000 MeV \\
\he4(\he3,$\gamma$)\be7    & 0.0000 MeV \\ \hline
\end{tabular}
\end{center}
\end{table}

\noindent
After some algebraic manipulation, we obtain the following evolution
equation for secondary production:
\begin{eqnarray}
\label{eq:secondary}
& & \left. \frac{d Y_S}{d z}\right|_{\rm sec} =
-\frac{rM_X\eta(n_\gamma^0)^2}{H_r \tau_X} (1+z)^3\exp{\left(
\frac{-1}{H_r\tau_X(1+z)^2} \right)}\sum_{T,{T^{^\prime}}} Y_T
Y_{T^{^\prime}} \int_0^\infty \!\!\!\!\! dE_A
\frac{\sigma_{A+T^{^\prime}\rightarrow S}(E_A) |v_A|}{b_A(E_A)}  
\nonumber \\
  & & \times
\int_{{\mathcal E}_A^{-1}(E_A)}^\infty \!\!\!\!\! dE_\gamma
\left(\frac{\tau_X}{E_0n_X}{\mathcal
N}_\gamma^{\rm QSE}(E_\gamma)\right) \sigma_{\gamma + T\rightarrow A}
(E_\gamma) \exp{\left[  -
\int_{E_A}^{{\mathcal E}_A(E_\gamma)} dE_A^{^{\prime \prime}}
\frac{\Gamma_A(E_A^{^{\prime \prime}})}{b_A(E_A^{^{\prime \prime}})}
\right] },
\end{eqnarray}
where $\eta$ is the baryon-to-photon ratio: $\eta \equiv n_{\rm 
B}^0/n_\gamma^0$.  

Table \ref{tab:nuke-thresh}
lists the secondary reactions considered. Deuterium production
does not occur within the lifetime range of interested to us, since the
neutron decays before it has a chance to react with a proton to form
deuterium, as pointed out in~\cite{ellis}.  Also, 
we have verified that the secondary production
of mass-7 elements is not significant, being small compared with that
produced during BBN.  The only significant secondary production is that of
\li6, as first shown in~\cite{kartsen} and later in~\cite{kkm}. We show
below that \li6 actually provides the strongest constraint for the
lifetime range we are interested in.
Note that the relevant threshold energies (Table \ref{tab:nuke-thresh})
 are those
in which the nuclei are in the cosmic rest frame,
and thus are computed in the fixed-target laboratory frame.  Consequently, 
these are about a factor of two higher than the center-of-mass
thresholds used in~\cite{kartsen} and~\cite{kkm}.

\section{Observational Constraints}

Before we discuss the results of our numerical analysis, we first discuss
the current status of the observational determinations of the
light-element abundances.  The abundances subsequent to any
photo-destruction/production must ultimately be related to these
observations.  Furthermore, our results are dependent on the assumed
baryon-to-photon ratio, which may either be determined through the
concordance of the BBN-produced abundances or through the analysis of the
CMB spectrum of anisotropies. As noted above, there is relatively good
agreement between the two.

\subsection{Observed Light Element Abundances}
\label{sect:obs}

Through painstaking observations of very different astronomical
environments, primordial abundances can be inferred for D, \he4, and \li7.
In addition, \he3 and \li6 have also been measured, and can provide
important supplementary constraints.  Here we summarize the data and our
adopted limits: more detailed reviews appear in~\cite{osw}. For all
nuclides, accurate abundance measurements are challenging to obtain, due
to systematic effects which arise from, e.g., an imperfect understanding
of the astrophysical settings in which the observations are made, and
from the process by which an abundance is inferred from an observed line
strength.

Deuterium is measured in high-redshift QSO absorption line systems via its
isotopic shift from hydrogen. In several absorbers of moderate column
density (Lyman-limit systems), D has been observed in multiple Lyman
transitions~\cite{omeara,tosl}. Restricting our attention to the three
most reliable regions~\cite{omeara}, we find a weighted mean of
\beq
\label{eq:D_p}
\pfrac{\rm D}{\rm H}_p = (2.9 \pm 0.3) \times 10^{-5}.
\eeq
We note, however, that the $\chi^2$ per degree of freedom is rather poor
( $\sim 3.4$), and that the unweighted dispersion of these data 
is $\sim 0.6\times 10^{-5}$. This
already points to the dominance of systematic effects. Observation of D in
systems with higher column density (damped systems) find lower
D/H~\cite{pb}, at a level inconsistent with \pref{eq:D_p}, further
suggesting that systematic effects dominate the error budget~\cite{foscv}.
If we used all five available observations, we would find D/H = $(2.6 \pm
0.3) \times 10^{-5}$ with an even worse $\chi^2$ per degree of freedom
($\sim 4.3$) and an unweighted dispersion of 0.8. As an upper limit to 
D/H, we
adopt the 2-$\sigma$ upper limit to the {\em highest} D/H value reliably
observed, which is D/H = $(4.0 \pm 0.65) \times 10^{-5}$, since we cannot
definitively exclude the possibility that some D/H destruction has
occurred in the other systems.

We also require a lower limit on the primordial D abundance.  Since
Galactic processes only destroy D, its present abundance in the
interstellar medium\cite{linsky}, ${\rm D/H} = (1.5 \pm 0.1) \times 
10^{-5}$ provides 
an extreme
lower limit on the primordial value, which is consistent with
(\ref{eq:D_p}). Therefore, we adopt the limits
\beq
\label{eq:D_lolim}
\EE{1.3}{-5} < \pfrac{\rm D}{\rm H}_p < \EE{5.3}{-5}.
\eeq
This lower bound is quite conservative, 
in light of the fact that the existence of heavy
elements confirms that stellar processing and thus D destruction has
certainly occurred at some level.

Unlike D, \he4 is made in stars, and thus co-produced with heavy elements.
Hence the best sites for determining the primordial \he4 abundance are in
metal-poor regions of hot, ionized gas in nearby external galaxies
(extragalactic H{\small II} regions). Helium indeed shows a linear
correlation with metallicity in these systems, and the extrapolation to
zero metallicity gives the primordial abundance (baryonic mass
fraction)~\cite{fo98}
\beq
Y_p = 0.238 \pm 0.002 \pm 0.005.
\label{he4}
\eeq
Here, the first error is statistical and reflects the large sample of
systems, whilst the second error is systematic and dominates.

The systematic uncertainties in these observations have not been
thoroughly explored to date~\cite{OSk}. In particular, there may be reason
to suspect that the above primordial abundance will be increased due to
effects such as underlying stellar absorption in the H{\small II} 
regions. We
note that other analyses give similar results: $Y_p = 0.244 \pm 0.002 \pm
0.005$~\cite{izotov} and 0.239 $\pm 0.002$~\cite{peim}.  For concreteness,
we use the \he4 abundance in (\ref{he4}) to obtain the range
\beq
\label{eq:he4-updown}
0.227 < Y_p < 0.249,
\eeq
taking the 2-$\sigma$ range with errors added in quadrature.

Helium-3 can be measured through its hyperfine emission in the radio band,
and has been observed in H{\small II} regions in our
Galaxy.  These observations find~\cite{bbr} that there are no obvious trends
in \he3 with metallicity and location in the Galaxy, but
rather a \he3 `plateau'. There is, however,
considerable scatter in the data by a factor $\sim 2$, some of which may
be real. Unfortunately, the stellar and Galactic evolution of \he3 is not
yet sufficiently well understood to confirm whether \he3 is increasing or
decreasing from its primordial value~\cite{vofc}.  Consequently, it is
unclear whether the observed \he3 `plateau' (if it is such) represents an
upper or lower limit to the primordial value. Therefore, we do not use
\he3 abundance as a constraint. If future observations of \he3 could
firmly establish the nature of its Galactic evolution, then \he3 could be
restored as a useful constraint on decaying particles, particularly in
concert with D.

The primordial \li7 abundance comes from measurements in the atmospheres
of primitive (Population II) stars in the stellar halo of our Galaxy. The
\li7/H abundance is found to be constant for stars with low metallicity,
indicating a primordial component, and a recent determination gives
\beq
\label{rfbon}
\pfrac{\li7}{\rm H}_p = \EE{(1.23 \pm 0.06_{-0.32}^{+0.68})}{-10}\ ({\rm
95\% \ CL}),
\label{firstLi}
\eeq
where the small statistical error is overshadowed by systematic
uncertainties~\cite{ryan}.  The range (\ref{rfbon}) may, however, be
underestimated, as a recent determination~\cite{liglob} uses a different
procedure to determine stellar atmosphere parameters, and gives
${\li7/{\rm H}}_p = (2.19 \pm 0.28) \times 10^{-10}$. At this stage, it
is not possible to determine which method of analysis is more accurate,
indicating the likelihood that the upper systematic uncertainty
in (\ref{rfbon}) has been underestimated.  Thus, in order to
obtain a conservative bound from
\li7, we take the lower bound (once again combining the statistical and
systematic errors in quadrature) from (\ref{rfbon}) and the upper bound
from~\cite{liglob}, giving
\beq
\label{eq:li7-updown}
\EE{9.0}{-11} < \pfrac{\li7}{\rm H}_p < \EE{2.8}{-10}.
\eeq

Finally, \li6 is also measured in halo stars, in which the \li6/\li7 ratio
is inferred from the (thermally blended) isotopic line splitting. The
lowest \li6 abudances comes from stars with primordial Li, which yield
$\li6/\li7 = 0.05 \pm 0.01$~\cite{li6obs}. The \li6 in these stars is not
primordial, as it is produced by cosmic-ray interactions with the
interstellar medium~\cite{li6cr}, predominantly $\alpha \alpha \rightarrow
\li6 + \cdots$. These same processes lead to the production of Be and B,
which are observed in halo stars at levels consistent with \li6 cosmic-ray
production. Since the observed \li6 abundances are consistent with being
entirely Galactic in origin, we can use these to set an extreme upper
limit on the primordial \li6 abundance. One complication enters, due to
the smaller binding energy of \li6 relative to \li7. This means that \li6
could in principle suffer depletion in stars due to nuclear burning,
without a similar depletion of \li7. However, once nuclear burning
becomes effective, \li6 depletion factors become extrelemly large making
such observations extremely unlikely \cite{bs}. It is therefore safe to
use the 2-$\sigma$ upper bound on the \li6/\li7 ratio
\beq
\label{eq:li67-updown}
\pfrac{\li6}{\li7} \la 0.07.
\eeq
Other depletion processes such as diffusion (included in the estimate of
systematic uncertainties in (\ref{rfbon})), would affect both \li6 and
\li7 similarly and not their ratio.  
It is also useful to consider the upper bound on \li6/H alone
\beq
\label{eq:li6-updown}
\pfrac{\li6}{\rm H}_p \la 2 \times 10^{-11}.
\eeq

\subsection{Cosmic Microwave Background Anisotropy Measurements}

Cosmic Microwave Background (CMB) anisotropy data are now reaching the
precision where they can provide an accurate measure of the cosmic baryon
content.  Given a CMB measurement of $\eta$, one can use BBN to make
definite predictions of the light element abundances, which can then be
compared with the observations discussed above. This comparison constrains
the effects of decaying particles more powerfully than if only the BBN
calculations were available to constrain $\eta$.

Recent results from DASI \cite{dasi} and CBI \cite{cbi}
indicate that $\Omega_B h^2 = 0.022^{+0.004}_{-0.003}$, while BOOMERanG-98
\cite{boom} gives $\Omega_B h^2 = 0.021^{+0.004}_{-0.003}$.
These determinations are somewhat lower than the central values found by 
MAXIMA-1
\cite{max}: $\Omega_B h^2 = 0.026^{+0.010}_{-0.006}$ and VSA
\cite{vsa}: $\Omega_B h^2 = 0.029 \pm 0.009$. Taking a
CMB value of
\beq
\Omega_B h^2 = 0.022 \pm 0.003 \qquad {\rm or} \qquad
\eta_{\rm 10,cmb} = 6.0 \pm 0.8
\label{etacmb}
\eeq
at the 1-$\sigma$ level, we would predict the
following light element abundances:
\begin{eqnarray}
\he4 : &\!\!& 0.248 \pm 0.001  \ (68 \% \ {\rm CL})\\
{\rm D/H} \times 10^5 : &\!\!& 2.7^{+0.9}_{-0.3} \ (68 \% \ {\rm CL})\\
{\rm \he3/H} \times 10^5 : &\!\!& 0.9 \pm 0.1 \ (68 \% \ {\rm CL})\\
{\rm \li7/H} \times 10^{10} : &\!\!& 3.4^{+1.5}_{-0.8} \ (68 \% \ {\rm
CL})
\end{eqnarray}
Note that these numbers are not outputs of BBN calculations corresponding
to $\eta_{10}=6.0$, but rather are the peak values of a likelihood
function found by convolving the results of the BBN Monte Carlo with an
assumed Gaussian for the distribution of CMB $\eta$ values. For further
details, see \cite{cfo1,cfo2}. With MAP  data, the accuracy of
$\eta_{\rm cmb}$ should be $10\%$ or better, which will give even tighter
predictions on the light elements.

\section{Model Results}
\label{sect:results}

We have implemented numerically the decaying-particle cascades discussed
in Section~\ref{sect:rxnet}. Using BBN light-element abundance 
predictions~\cite{cfo1}
as initial conditions, we calculate the final abundances for
particular sets of baryon and dark matter parameters. The
three free
parameters are: 
\beq
\zeta_X \equiv  \frac{n_X^0}{n_\gamma^0} M_X
=  rM_X = 2rE_0 ,
\eeq
$\tau_X$ and $\eta$. 

\subsection{Analytic Discussion}

Some simple analytic approximations allow us to gain insight into the
essential physics in our problem. As we will see, the following analytic
treatment reproduces well the behavior of the light element abundance
mountains and deserts in our parameter space.

The dependence on $\tau_X$ can be understood~\cite{holtmann} in
terms of the characteristic energy scales in the photon spectrum
(\ref{eq:injection}). Both the break $E_X$ and the cutoff $E_C$ scale as
$E_i \propto 1/T$. Thus, in the `uniform decay' approximation where all
particles decay at $t = \tau_X$, the decay occurs at $T \sim 10^{-4} \,
{\rm MeV} \, (\tau_X/10^8 \, {\rm s})^{-1/2}$. Consequently, we have $E_X
\sim 28 \, {\rm MeV} \, (\tau_X/10^8 \, {\rm s})^{1/2}$, and $E_C \sim 103
\, {\rm MeV} \, (\tau_X/10^8 \, {\rm s})^{1/2}$, and cutoffs thus increase
with $\tau_X$. In other words, higher-energy photo-erosion processes can
occur for longer lifetime values. Comparison with
Table~\ref{tab:photo-thresh} shows that, as $\tau_X$ increases, first
$E_C$ and then $E_X$ pass the threshold energy $E_{\rm th}$ for any given
process, at which point the process becomes important. A reaction can
turn on when $E_C \ga E_{\rm th}$, which in the uniform decay
approximation occurs when
\beq
\label{eq:dropout}
\tau_X \ga 10^6 \ {\rm s} \ \pfrac{E_{\rm th}}{10 \ {\rm MeV}}^2,
\eeq
while for shorter $\tau_X$ the channel is closed.
The reaction grows stronger when $E_X \ga E_{\rm th}$, 
which occurs when 
\beq
\label{eq:tauX}
\tau_X \sim 10^7 \ {\rm s} \ \pfrac{E_{\rm th}}{10 \ {\rm MeV}}^2.
\eeq
We can also understand the $\zeta_X$ dependence of photodestruction (and
secondary production) analytically, as follows. In the limit of small
$\zeta_X$, the decaying particle has no influence on the light-element
abundances as predicted by primordial nucleosynthesis, predicting a
universe made of mostly hydrogen and \he4, with small but significant
amounts of D, \he3, and \li7.  Lithium-6 is not produced in significant
quantities.  Going beyond this trivial case we use a similar
treatment as above, and employ the uniform decay approximation. To begin, 
as long as a reaction can proceed, a typical shower photon has energy
\beq
\langle E \rangle = 56\ {\rm MeV} \pfrac{E_{th}}{10\ {\rm
MeV}}^\frac{1}{2} \pfrac{\tau_X}{10^8\ {\rm s}}^\frac{1}{4}, 
\eeq
so that the number of such photons per decay is $N_\gamma \sim E_0/\langle
E\rangle$. Had the lower-energy piece of the power law been much steeper
(i.e., with a power index $p>2$) we would have $\langle E\rangle \sim 
E_{th}$, and if it
was much shallower (i.e., $p<1$) we would have $\langle E\rangle \sim 
E_X$.  However, we
lie in an interesting regime where $\langle E\rangle \sim
(E_{th}/E_X)^{P-1} E_X$, where $1<p<2$.  Thus the nonthermal photon
density is

\beq
n_{\gamma}^{\rm inj} = N_\gamma n_X = \frac{\zeta_Xn_\gamma^{BG}}{2\langle 
E\rangle}.
\eeq
These photons are thermalized at a rate per photon of $\Gamma_{\rm therm} \sim
n_e \sigma_{\rm T} \sim n_{\rm B} \sigma_{\rm T} $. The rate per photon
for the photodestruction of species $T$ to yield species $P$ is
$\Gamma_{T\rightarrow P} \sim n_T \sigma_{T\rightarrow P} =
f_{T\rightarrow P} Y_T \Gamma_{\rm therm}$, where $Y_T = n_T/n_{\rm B}$, and
$f_{T\rightarrow P} = \sigma_{T\rightarrow P}/\sigma_{\rm T}$ is the
relative strength of the cross section for photodestruction of $T$ into
$P$, compared with the thermalization cross section, which we take as the
Thompson cross section for this discussion.  Consequently, the change in
the number density of $A$ is the net production rate per volume
$(\Gamma_{T\rightarrow A} - \Gamma_{A\rightarrow P}) n_\gamma^{\rm inj}$ times
the loss time $\Gamma_{\rm therm}^{-1}$, or
\beq
\delta n_A = n_B \delta Y_A \sim n_\gamma^{\rm inj}\frac{(\Gamma_{T\rightarrow A}
- \Gamma_{A\rightarrow P})}{\Gamma_{\rm therm}}.
\eeq
We see that the fractional change in an abundance is given by
\beq
\frac{\delta Y_A}{Y_A} \sim \frac{\zeta_X}{2\eta\langle E\rangle}\left(
\frac{Y_T}{Y_A} f_{T\rightarrow A} - f_{A\rightarrow P}\right).
\eeq
If we look at the two extremes when either production or destruction
dominates, we can derive the behavior of $\zeta_X$, given that the
fractional change in the abundance is $\sim 1/2$:
\beqar
\label{eq:prod}
\zeta_X^{\rm prod} &\sim& \EE{3.2}{-11}\ {\rm GeV} \left( 5000 \over
\frac{Y_T}{Y_A} \right) \left( \EE{5.0}{-4} \over f_{T\rightarrow A}
\right) \left( \eta_{10} \over 6 \right) \left( E_{th} \over 20{\ \rm
MeV} \right)^\frac{1}{2} \left( \tau_X \over 10^8{\ \rm s}
\right)^\frac{1}{4}, \\
\label{eq:dest}
\zeta_X^{\rm dest} &\sim& \EE{6.3}{-8}\ {\rm GeV} \left( \EE{5.0}{-4}
\over f_{A\rightarrow P} \right) \left( \eta_{10} \over 6 \right)
\left( E_{th} \over 2.224{\ \rm
MeV} \right)^\frac{1}{2} \left( \tau_X
\over 10^8{\ \rm s} \right)^\frac{1}{4},
\eeqar
where the numbers are those appropriate for D.  

This same treatment can be extended to secondary production of light
elements. Since \li6 is the only significant secondary production, this 
is the only example we consider here. In this case, only a fraction of
the primary products have enough energy for this reaction to proceed,
because of interactions with the background plasma.  We estimate as
follows the fraction that can react to form \li6.  Each prospective
reactant is produced with initial energy ${\mathcal E}_A(\langle E\rangle
)$, and the total amount of energy that can be lost between collisions is
${\mathcal E}_{loss} = b_A(\langle E\rangle )/n_{\rm B}\sigma_T|v_A|$.  
The fraction of reactants left is the ratio of these two energies, $N_P
\sim {\mathcal E}_A(\langle E\rangle )/{\mathcal E}_{loss} \sim 0.001$.  
Thus the number density of these remaining non-thermal particles is
\beq
n_P^{\rm inj} = N_P\delta n_A \sim N_P\frac{\zeta_XY_Tn_{\rm B}}{2\eta \langle
E\rangle }f_{T\rightarrow A}.
\eeq
The rate per particle can be described in a similar way as before, where
$\Gamma_{A\rightarrow S} = f_{A\rightarrow S}Y_{T^{^\prime}}\Gamma_{\rm therm}
\beta$.  Since these particles are non-relativistic, there is a factor
$\beta = v/c$ in the interaction rate.  We can thus determine the change
in abundance of the secondary species:
\beq
\delta n_S \sim n_{\rm B} \delta Y_S \sim n_P^{\rm inj}\frac{\Gamma_{A\rightarrow
S}}{\Gamma_{\rm therm}}.
\eeq
The fractional change in the abundance is then given by
\beq
\frac{\delta Y_S}{Y_S} \sim N_P
\frac{\zeta_XY_TY_{T^{^\prime}}}{2\eta \langle E\rangle Y_S}
f_{T\rightarrow A}f_{A\rightarrow S}\beta.
\eeq
Using parameters appropriate for \li6 and $\beta \sim 0.01$, we derive
the value of $\zeta_X$ when secondary production becomes important:
\beq
\label{eq:sec}
\zeta_X^{sec} \sim \EE{2.6}{-12}{\ \rm GeV}
\pfrac{\EE{2.5}{11}}{\frac{Y_TY_{T^{^\prime}}}{Y_S}}
\pfrac{\EE{5.0}{-4}}{f_{T\rightarrow A}}
\pfrac{0.05}{f_{A\rightarrow S}} \pfrac{\eta_{10}}{6}
\pfrac{E_{th}}{20{\ \rm MeV}}^\frac{1}{2} \pfrac{\tau_X}{10^8{\ \rm
s}}^\frac{1}{4}.  
\eeq
One should note that, at the high photon energies required to induce
these tertiary reactions, double-photon scattering is comparable to the 
Compton scattering and nuclear pair-production mechanisms for photon 
energy-loss. This weakens the dependence of the reaction rates on the
baryon density.


We now turn to the full numerical results, using these analytical
estimates as a guide to interpretation.

\subsection{Numerical Results}

\begin{figure}
\epsfig{file=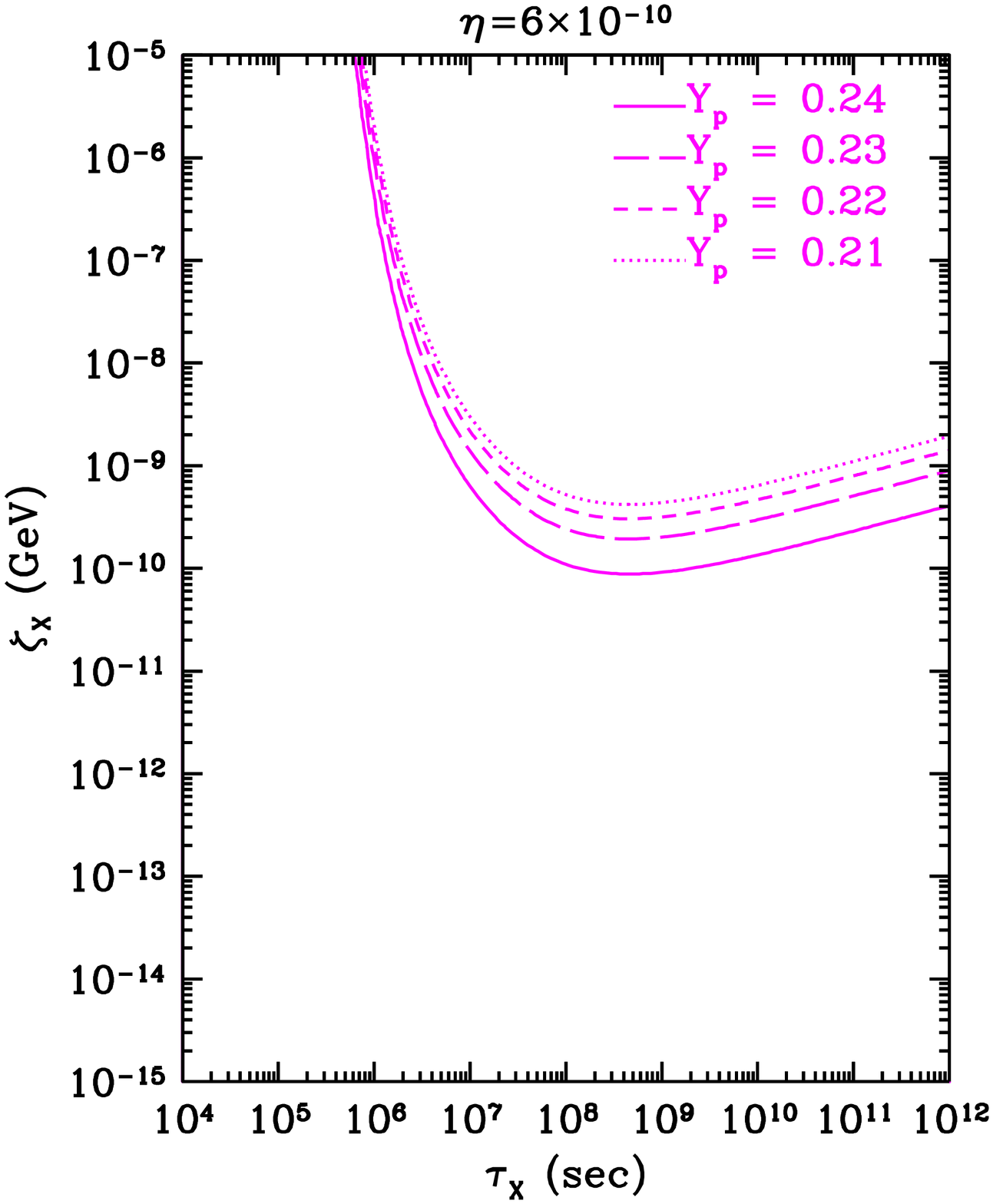,height=4in}
\hfill
\epsfig{file=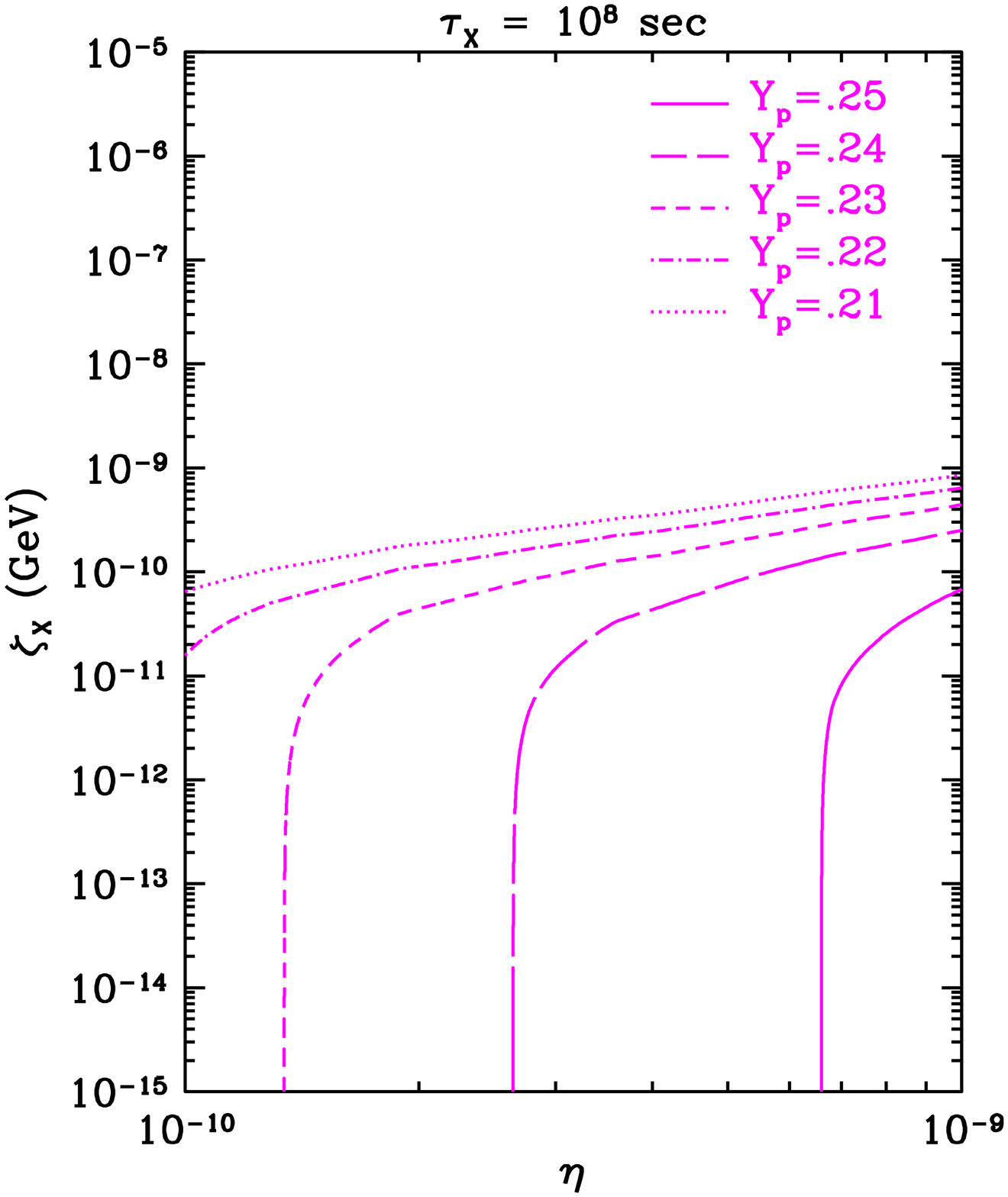,height=4in}
\vskip -.3in
\caption{\it
Contours of the \he4 mass fraction $Y_p$
(a) in the $(\zeta_X, \tau_X)$ plane, 
for  $\eta_{10} = 6$, and (b) in the $(\zeta_X, \eta)$ plane,
for $\tau_X = 10^8$ sec.
See the text for discussion.
}
\label{fig:He4tau}
\end{figure}

Our numerical results
come from the integration of \pref{eq:primary} and \pref{eq:secondary},
using the input spectra of \pref{eq:injection}.
We present our results by showing abundance contours
in $\zeta_X-\tau_X$ space at fixed $\eta_{10}=6$, and
in $\zeta_X-\eta$ space at fixed $\tau_X = 10^8$~s.
A key innovation of the present work is a detailed fitting of
the energy dependence of the
relevant cross sections.  These are discussed in Appendix A,
which also contains fitting formulae.

We begin our discussion with the most abundant compound nucleus, namely
\he4.  Since the other light elements are predicted by BBN to have
abundances that are orders of magnitude smaller than \he4, no significant
production of \he4 can take place. Thus electromagnetic showers from
decaying particles can only destroy \he4.  These photodestruction
processes have an energy threshold of $E_{\rm th} \sim 20$ MeV and so,
from (\ref{eq:dropout}), we expect this process to become inefficient for
$\tau_X \la 4\times 10^6$ s, and to shut down completely when $\tau_X \la
4 \times 10^5$ seconds. Indeed, this is what is seen in
Fig.~\ref{fig:He4tau}(a), where we plot contours of $Y_p$ in the
$(\zeta_X, \tau_X)$ plane for $\eta_{10}=6$, the value preferred by CMB
analyses (\ref{etacmb}). For $\tau_X \ga 4 \times 10^6$~s, the \he4
destruction factor goes from a small perturbation to a large one as
$\zeta_X$ grows from $10^{-10}$ to $10^{-9}$~GeV, until the region
$\zeta_X \ga 10^{-8}$~GeV becomes a \he4 `desert'. Over the region $\tau_X
\sim 4 \times 10^{5} \, {\rm s} - 4 \times 10^{6} \, {\rm s}$, \he4
destruction becomes important only at increasingly high $\zeta_X$.  This
general behavior has an impact on all of the other light elements, as \he4
is the only important source for them.

In Fig.~\ref{fig:He4tau}(b), we again plot contours of Y$_p$, but now in
the $(\zeta_X,\eta)$ plane for $\tau_X = 10^8$~s.  We see the generic
features mentioned above, that for low $\zeta_X$ \he4 is at its BBN
predicted value and for very large $\zeta_X$ it is destroyed. At large
$\eta$, the non-thermal photons are more quickly thermalized, thus having
less energy on average and making it more difficult to destroy nuclei.  
This linear rise is as predicted in \pref{eq:prod}.


\begin{figure}
\epsfig{file=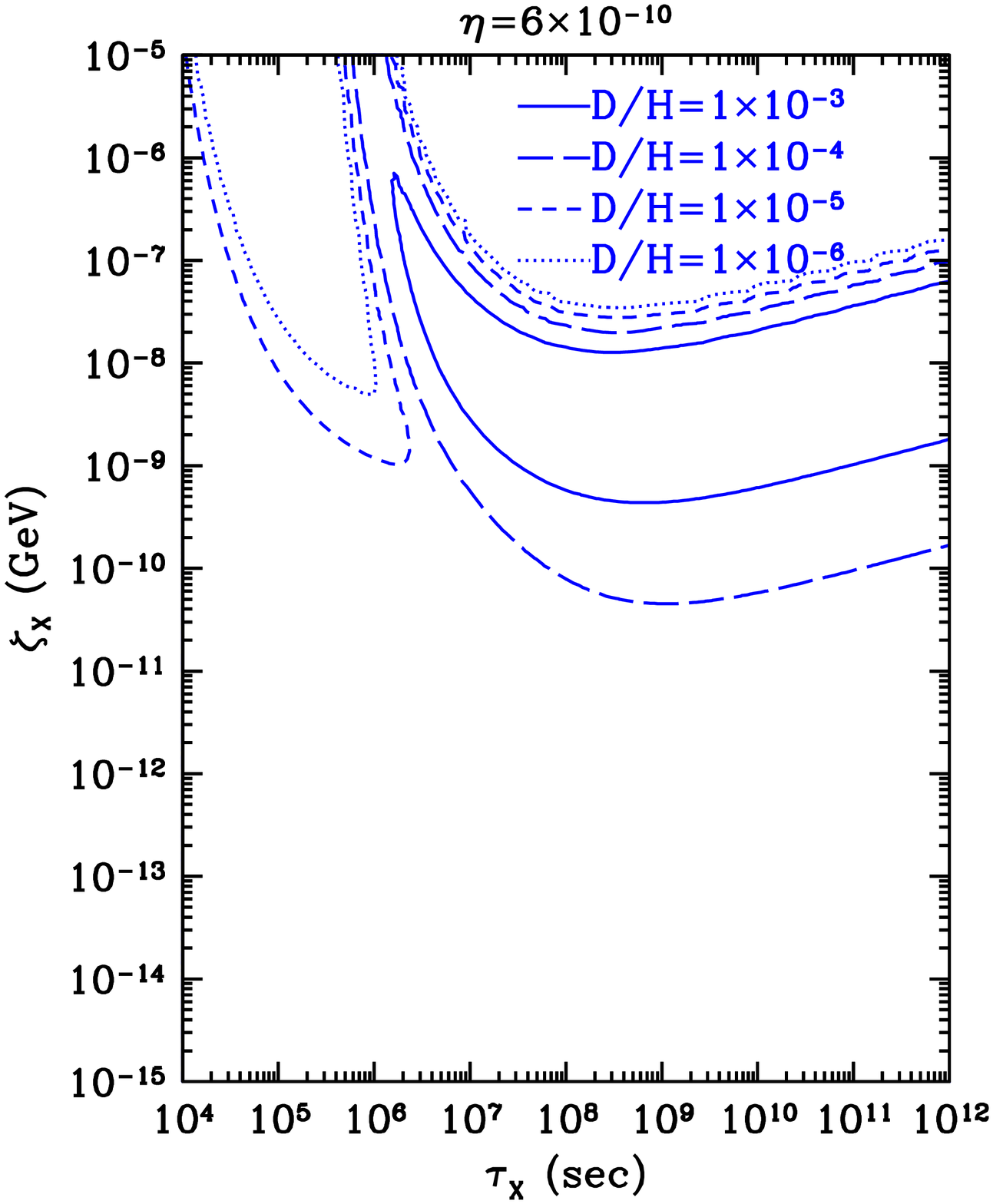,height=4in}
\hfill
\epsfig{file=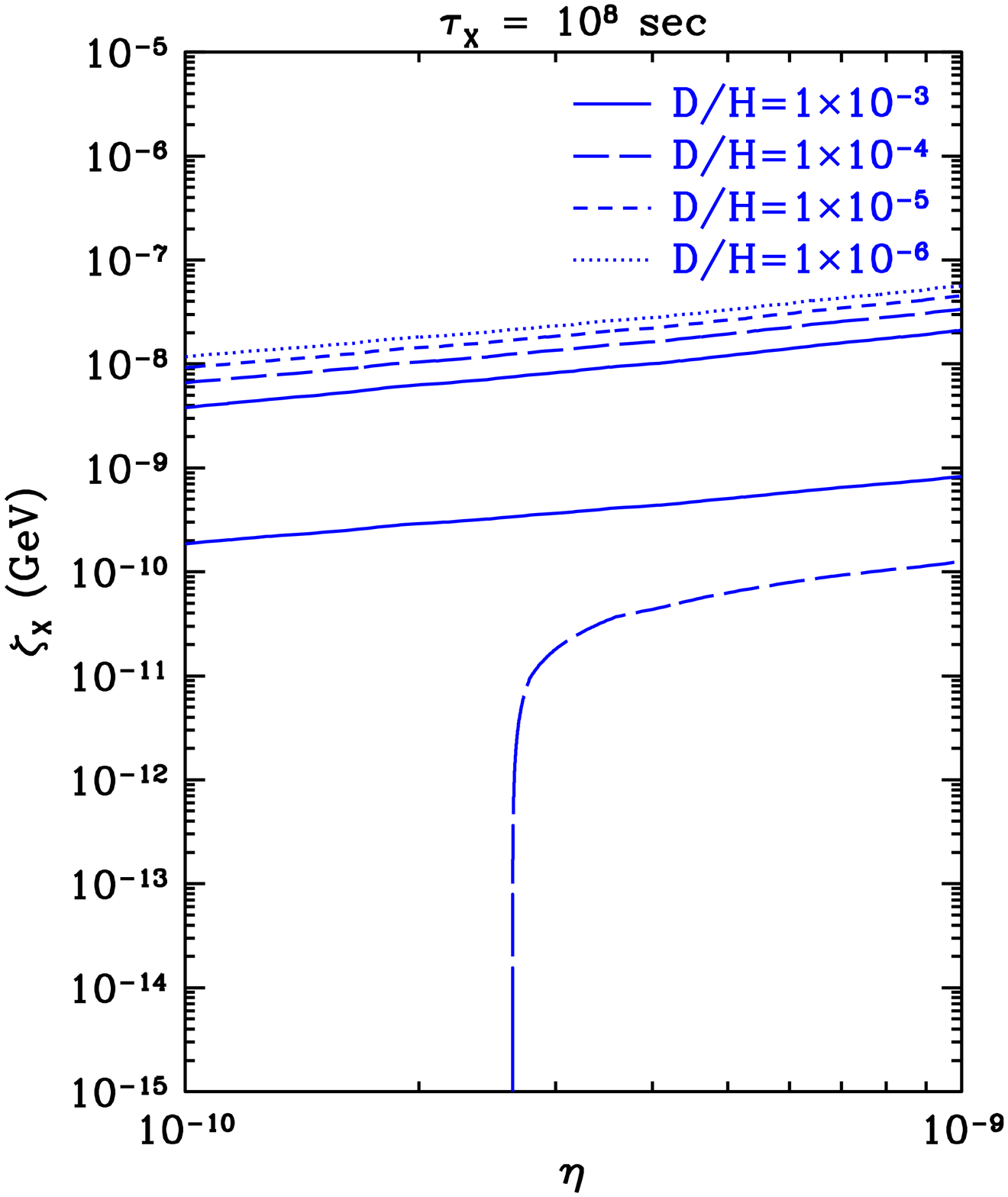,height=4in}
\caption{\it
Deuterium abundance contours 
plotted as in Fig.~\ref{fig:He4tau}.
}
\label{fig:Dtau}
\end{figure}

Fig.~\ref{fig:Dtau} plots the corresponding contours of D/H. Several
distinct regions are apparent. For all $\tau_X$, in the region $\zeta_X
\la 10^{-11}$~GeV, the decaying particles make only a small perturbation
to the primordial value of D/H.  For values of $\zeta_X \ga 10^{-11}$~GeV,
the decaying particles can lead to significant perturbations to D/H, the
sign and magnitude of which depend strongly on $\tau_X$.  In
the case of D, the important production processes are due to $\he4 +
\gamma$: the other light elements have negligible abundances compared to
\he4, and thus are unimportant as D sources. The D production channels
have thresholds $E_{\rm th} \sim 25$~MeV, so we expect these process to
become inefficient for $\tau_X \la 5 \times 10^6$~s.  Furthermore,
production is only efficient when \he4 destruction is not large, i.e.,
when $\zeta_X \la 10^{-9}$~GeV. Of course, D production can only occur
when there is sufficient \he4 destruction, so we expect significant
production to occur only up to some maximum $\zeta_X$.  For higher
$\zeta_X$, photo-destruction becomes so dominant that any production
yields are immediately broken up via further photo-destruction reactions,
leaving a universe filled with only protons. Thus, as $\zeta_X$ increases,
the D/H abundance rapidly declines, dropping to and then below its
primordial abundance, to approach zero in a D `desert'. Thus, we expect D
production only in a region bounded from above and below in $\zeta_X$, and
to the left by $\tau_X \la 4 \times 10^6$~s.  These expectations are met
by the D/H `mountain' in Fig.~\ref{fig:Dtau}, which stretches between
$\zeta_X = 10^{-10}$~GeV to $\zeta_X = 10^{-8}$~GeV.

For $\tau_X \la 5 \times 10^6$~s, D production from \he4 does not occur.
Since the photo-destruction processes $d(\gamma,n)p$ has a threshold of
$E_{\rm th} = 2.224$~MeV (the D binding energy), D destruction drops out
at $\tau_X \la 4 \times 10^3$~s, as seen in Fig.~\ref{fig:Dtau}. Finally,
note that the competing processes of D production and destruction balance
for some regions of parameter space, where D retains its primordial
values. These are the `channels' which separate the regions we have
already discussed.

In Fig.~\ref{fig:Dtau}(b), the contours of D/H are shown in the
$(\zeta_X,\eta)$ plane for $\tau_X=10^8$~s.  Just like \he4, these D
contours show the features sketched out in our analytic discussion.  For
low $\zeta_X$, D is at its BBN value.  At intermediate values, we are
climbing the photo-production mountain.  At even higher values of
$\zeta_X$, we enter the photo-erosion desert.  Again the dependences on
$\eta$ are entirely due to the photon energy-loss mechanism being more
efficient at higher baryon density.


\begin{figure}
\epsfig{file=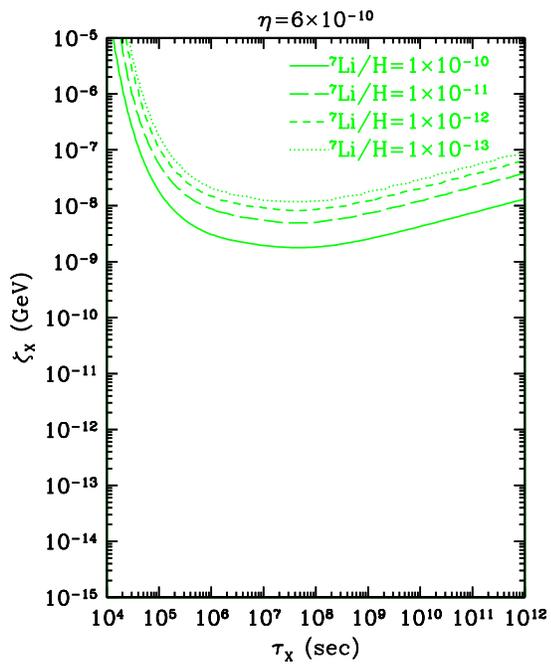,height=4in}
\hfill
\epsfig{file=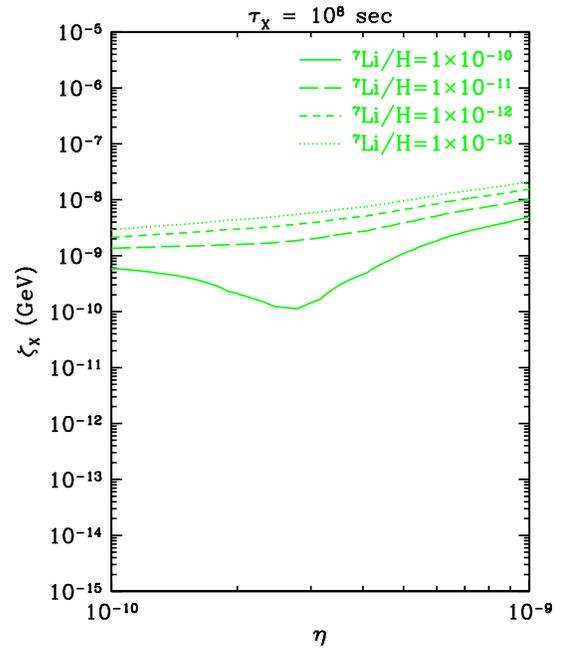,height=4in}
\caption{\it
Contours of \li7/H, plotted
as in Fig.~\ref{fig:He4tau}.
}
\label{fig:Li7tau}
\end{figure}

Results for \li7/H are shown in Fig.~\ref{fig:Li7tau}. We see that the
results are qualitatively similar to those for \he4, reflecting the fact
that secondary \li7 production is negligible.  Since \li7 is only
destroyed, its weak binding compared to \he4 leads to the wider expanse of
the \li7 `desert'.  As mentioned earlier, we considered the secondary
production of \li7, through the reactions \he4(t,$\gamma$)\li7 and
\he4(\he3,$\gamma$)\be7.  These reactions have no energy threshold, only a
strong Coulomb barrier, so {\it a priori} they would seem important.  
However, the net production of \li7 through these reactions is small
compared to its primordial value set by BBN. This is due to the small
cross sections for mass-7 production.


\begin{figure}
\epsfig{file=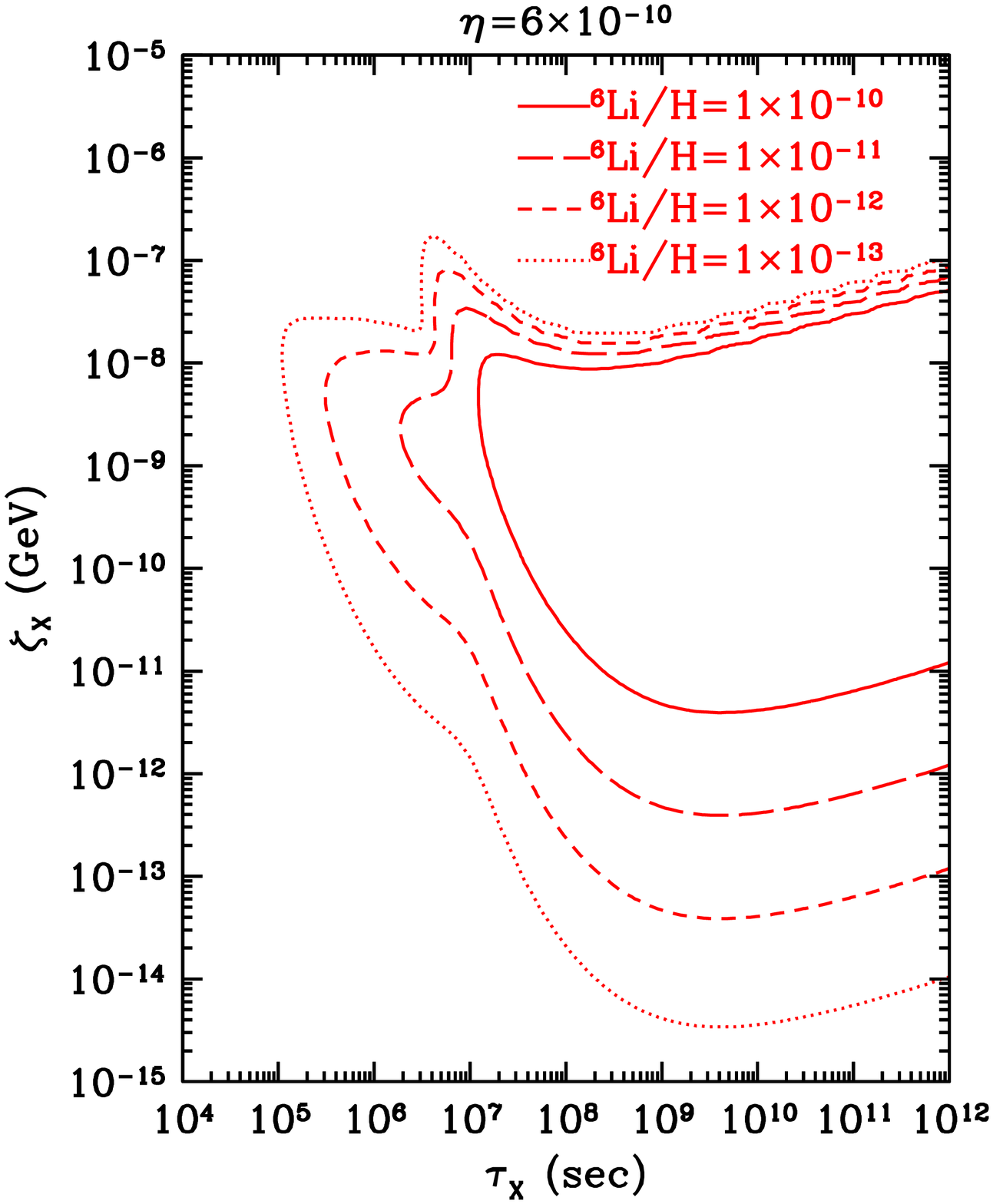,height=4in}
\hfill
\epsfig{file=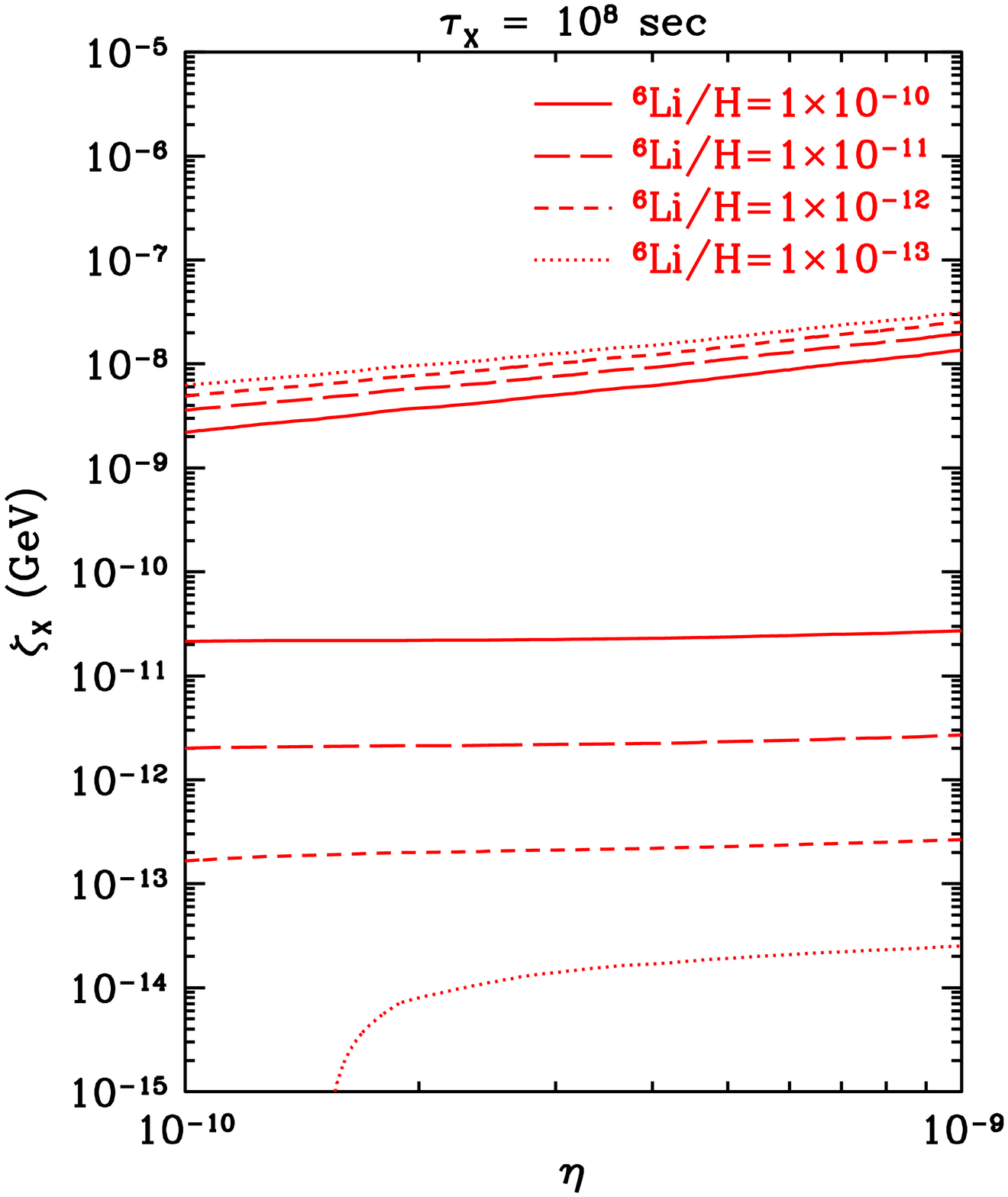,height=4in}
\caption{\it
Contours of \li6/H, plotted
as in Fig.~\ref{fig:He4tau}.
}
\label{fig:Li6tau}
\end{figure}

Unlike \li7, standard BBN does not
produce \li6 in any observable quantity, so any other production mechanism
is important.  
The behavior of \li6, seen in Fig.~\ref{fig:Li6tau}a, can be understood in
terms of the \he4 and \li7 dynamics, because these are the two \li6
sources. As with D, we see a roughly horizontal `mountain' of \li6
production, which is bounded on the left by threshold effects. The
dominant production channel is from secondary reactions, and thus is tied
to the \he4 destruction threshold.  The secondary reactions \he4(t,n)\li6 and \he4(\he3,p)\li6,
also have low energy resonances, further increasing their yields.  The
\li6 production from \li7 and \be7, however, has lower thresholds, and
thus becomes dominant for small lifetimes, $\tau_X \la 10^7$~s. These
channels are important where \li7 destruction is moderate but still
sufficient to make significant amounts of \li6, leading to a break in the
slope of the \li6 curve at $\zeta_X \sim 10^{-11}$. The inclusion of two
\li6 destruction channels leads to the `nose' at $\zeta_X \sim
10^{-7}$~GeV and $\tau_X \sim 10^{7}$~s. This feature is absent in the
\li6 plot in~\cite{kkm}, since there only a single destruction channel was
considered.

In Fig.~\ref{fig:Li6tau}b we see that the secondary production dominates
the evolution of \li6 for quite low $\zeta_X$.  Since the \li6
secondary-production cross sections have thresholds, the average
initiating photon energy must be higher than in the standard
photo-destruction process.  In this higher-energy regime, double-photon
scattering is comparable to the other photon energy-loss mechanisms.  This
reduces the dependence on the baryon density, effectively flattening out
the contours in the $(\zeta_X,\eta)$ plane where secondary production is
important.  At higher $\zeta_X$, photo-destruction of \li6 takes over,
taking us to the \li6 desert.

\begin{figure}
\epsfig{file=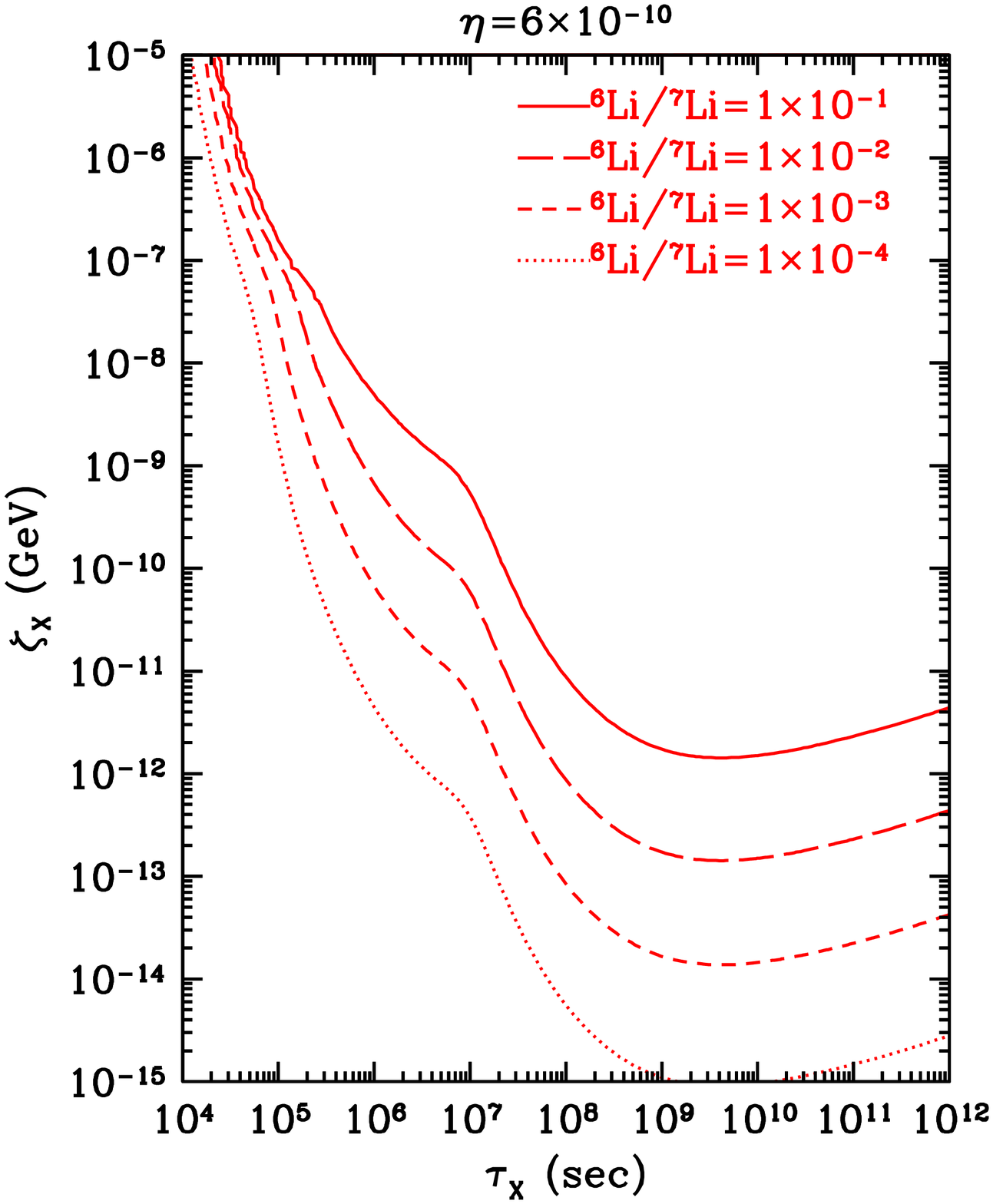,height=4in}
\hfill
\epsfig{file=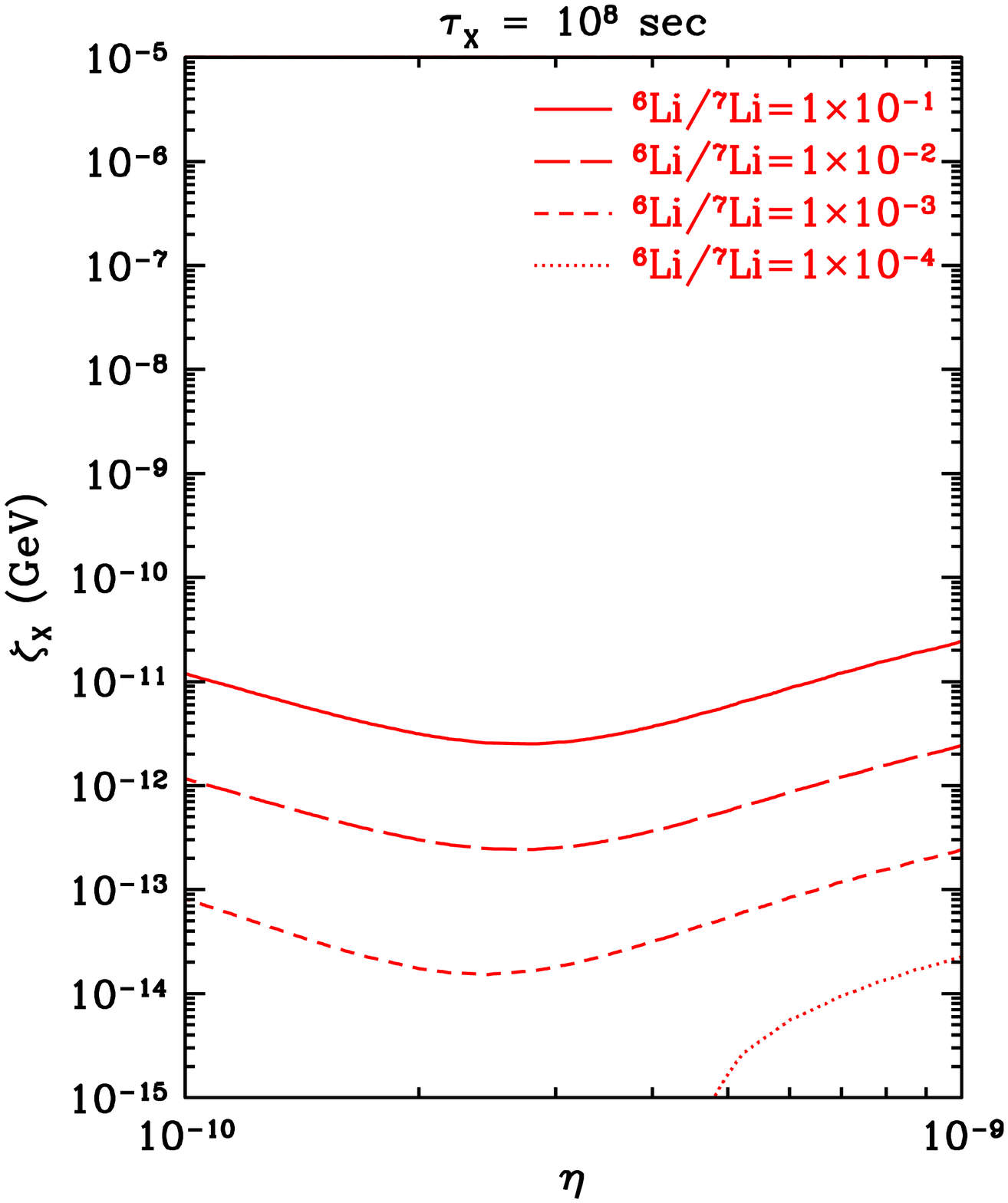,height=4in}
\caption{\it
Contours of \li6/\li7, plotted
as in Fig.~\ref{fig:He4tau}.
}
\label{fig:67ratio}
\end{figure}

As noted in~\cite{holtmann}, the \li6/\li7 ratio offers additional
constraints besides those provided by each nuclide separately. We plot
\li6/\li7 contours in Fig.~\ref{fig:67ratio}, where we see that these are
smoother than the contours for \li6 alone. Because \li6 dissociation has a
higher threshold than \li7 - see Table~\ref{tab:photo-thresh} - \li6/\li7
is large in the `desert' region of high $\zeta_X$, though each individual
abundance is quite small. For smaller $\zeta_X$, \li6 increases with
$\zeta_X$, while \li7 either remains constant or decreases.  The upshot is
that \li6/\li7 grows with $\zeta_X$, as seen in Fig.~\ref{fig:67ratio}.

Having described the physics that leads to the abundance patterns we have
computed, we can now discuss how this physics allows the observed
abundances to place constraints on decaying particles. To obtain these
constraints, we combine this analysis with the observational data
discussed previously.

\section{Limits on Unstable Relic Particles}
\label{sect:limits}

We now impose the observed light-element abundance constraints of Section
\ref{sect:obs} with the results of the previous section.  We do this for
each element individually, then combine the results to obtain the
strongest constraints.

We remind the reader that light element constraints
on decaying particles depend on $\eta$, but of course
one cannot use the standard BBN limits on $\eta$ as
part of ones limits.  In  the past, this difficulty has
only been overcome by adopting limits on $\eta$
derived from non-nucleosynthetic arguments.
These limits have, until recently, been rather weak, 
which has weakened the power of the light element
constraints.  This situation has now changed drastically.
We recall that the CMB now imposes $\eta_{10} \simeq 6$
with rather small uncertainty.
Thus, if we adopt the CMB results, we no longer
must treat $\eta$ as weakly constrained by
non-BBN arguments, strengthening the results we derive.

We quote limits for $\eta_{10} = 6$,
and emphasize results for $\tau_X = 10^8$~s, 
which is roughly the lifetime for which the constraints
are the strongest, and is also within
the range of current interest for gravitino decays.
Results at lower $\tau_X$ weaken rapidly, while 
the constraints at higher $\tau_X$
scale roughly as $\tau_X^{1/4}$, as in
\pref{eq:prod}, \pref{eq:dest} and \pref{eq:sec}.

Also, the current observational status of standard BBN comes into play.  
Namely, the present observational data on \he4 and \li7 are in tension
with those for D, the former preferring $\eta_{10} \sim 3$ and the latter
$\eta_{10} \sim 6$. On the one hand, if standard BBN is correct and the
tension is due to systematic errors, the result is that these errors
weaken the constraints that one can place on decaying particles. On the
other hand, if the observations were to improve to the point that the
light-element disagreement can no longer be accommodated, this could herald
new physics.  In this case, decaying particles offer a way~\cite{holtmann}
of reconciling the abundances of the light elements, in which case one
also derives estimates of the required $\zeta_X$ and $\tau_X$.

We turn first to the elements which are only
destroyed, namely \he4 and \li7.
The observed constraints on \he4 \pref{eq:he4-updown}
give
\beq
\zeta_X(\he4) < \EE{2.5}{-10} \ {\rm GeV},
\eeq
which is driven by the lower limit
$Y_p > 0.227$ that we adopted.
The sharp drop of \he4 with increasing
$\zeta_X$ (i.e., the descent into the desert of Fig.
\ref{fig:He4tau})
ensures that the constraint on $\zeta_X$ is
insensitive to the precise $Y_p$ limit chosen.
The situation is similar for \li7,
for which  \pref{eq:li7-updown}
give the weaker constraint
\beq
\zeta_X(\li7) < \EE{2}{-9} \ {\rm GeV}.
\eeq

For deuterium, net production and net destruction are
both possible.  In terms of Fig. \ref{fig:Dtau},
this means that limits on D exclude the
ridge in the D mountain, while allowing
regions at higher and lower $\zeta_X$.
In particular, the observed D abundances \pref{eq:D_p}
allow the
range 
$\EE{2}{-8} \ {\rm GeV} \la \zeta_X \la \EE{3}{-8} \ {\rm GeV}$,
but the \he4 and \li7 constraints are each able to
exclude this regime.
Consequently, the only remaining
region is the low-$\zeta$ side of the mountain,
\beq
\zeta_X({\rm D}) < \EE{3}{-11} \ {\rm GeV}.
\eeq

The since \li6 is not produced significantly in standard BBN,
only production is important for low $\zeta_X$, while
for higher $\zeta_X$ destruction dominates.  
Thus, the situation is similar to that of D:
there is a \li6 mountain, with the observations allowing
a narrow high-$\zeta_X$ region and a large low-$\zeta_X$ region.
The \li6/H abundance of \pref{eq:li6-updown}
gives 
\beq
\zeta_X(\li6) < \EE{5}{-12} \ {\rm GeV}
\eeq
in addition to a higher region that is discordant with
\he4 and \li7.
The 
\li6/\li7 ratio
\pref{eq:li67-updown}
gives 
\beq
\zeta_X(\li6/\li7) < \EE{7.0}{-12} \ {\rm GeV}.
\eeq

\begin{figure}
\psfig{file=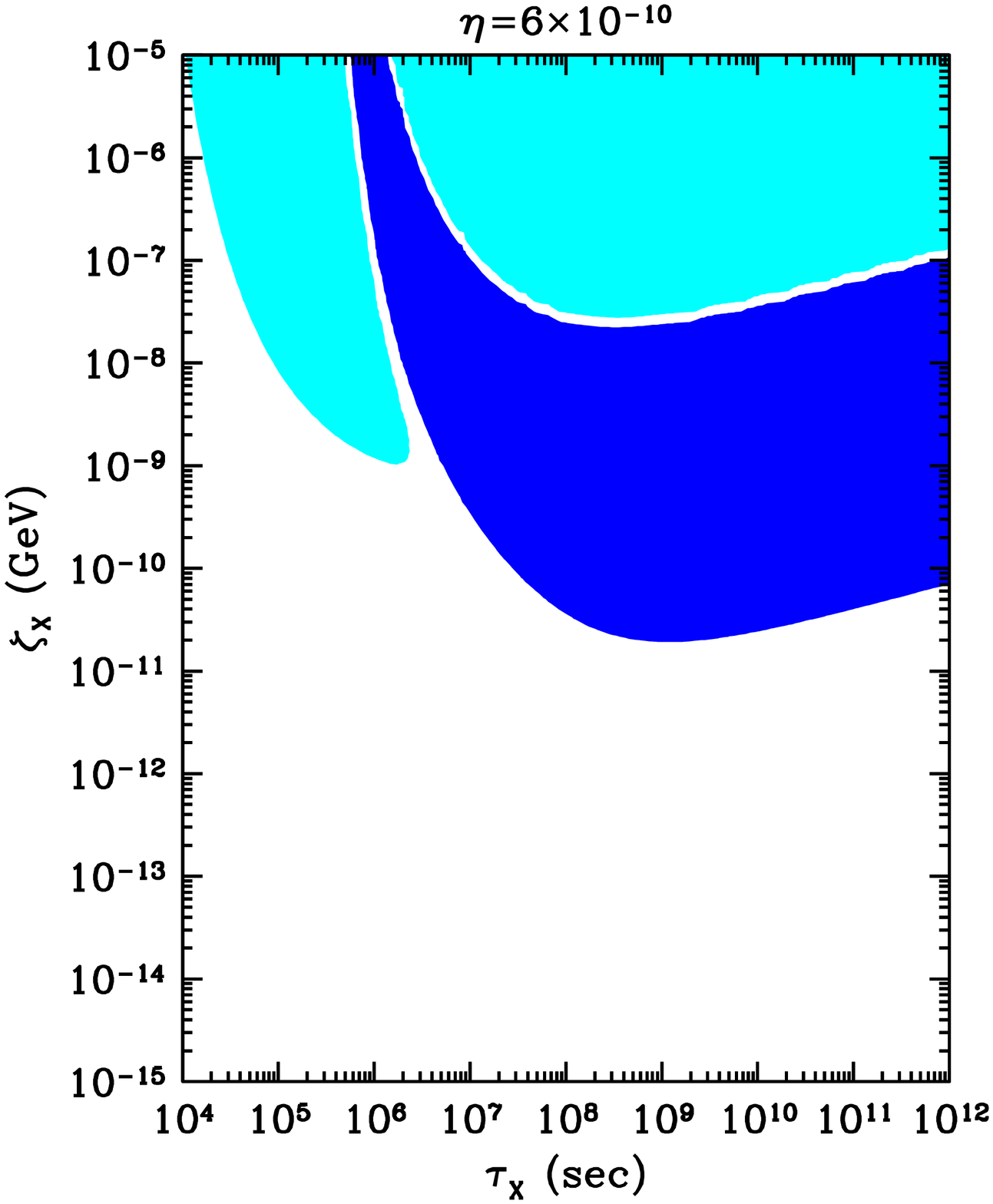,height=4in}
\hfill
\psfig{file=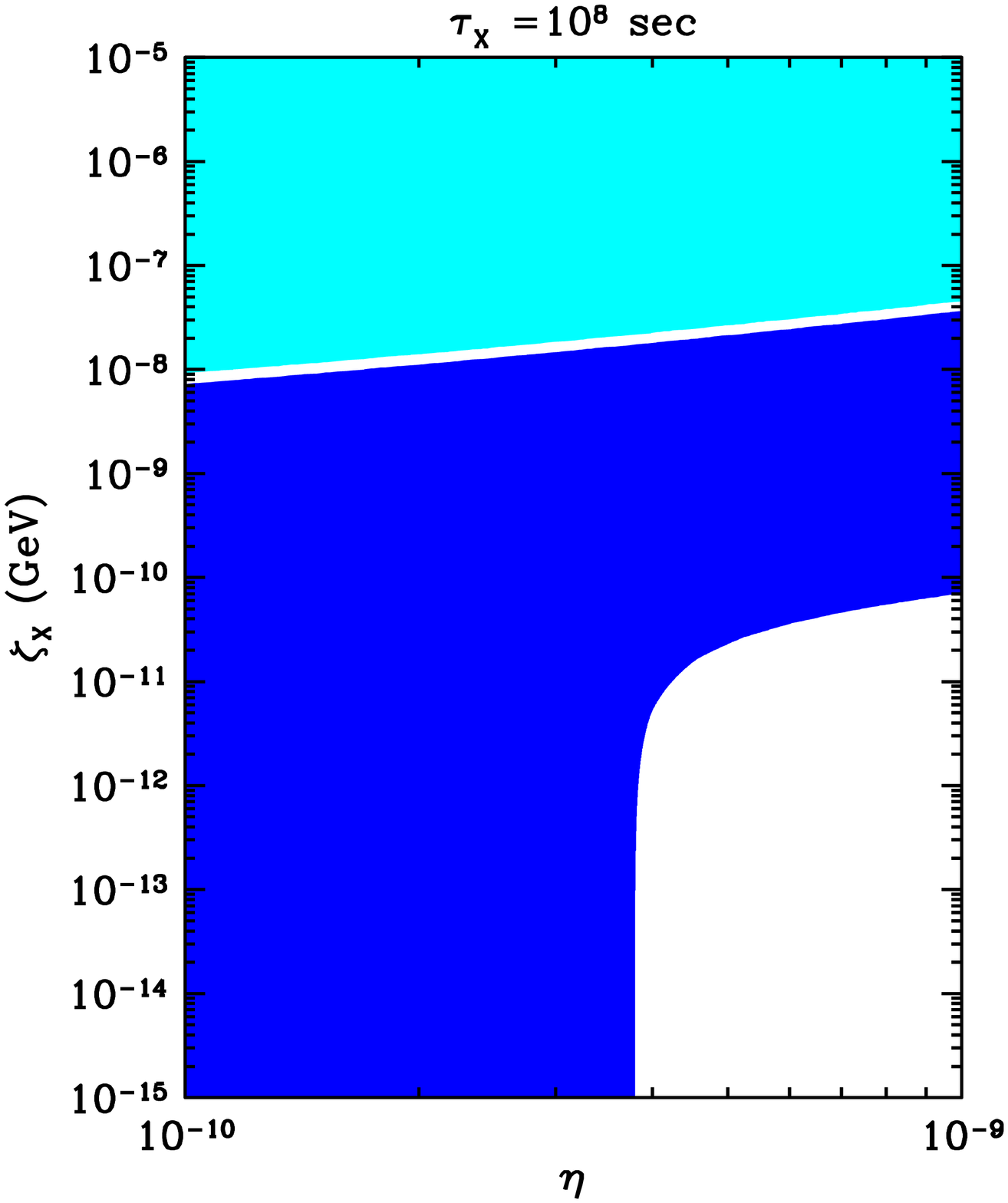,height=4in}
\caption{\it
Exclusion limits based on deuterium are shown in  (a) the 
$(\zeta_X,\tau_X)$ plane for $\eta_{10}=6$ (b) the $(\zeta_X,\eta)$ plane
for $\tau_X=10^8$~s. The dark (blue) shaded region corresponds to an 
overabundance of D/H, while the light (blue) shaded region corresponds to an
underabundance of D/H.
 } 
\label{fig:combined} 
\end{figure}

Figure \ref{fig:combined} summarizes our results for the constraints based
on D/H in both the $(\zeta_X,\tau_X)$ plane (for $\eta_{10}=6$) and 
the $(\zeta_X,\eta)$ plane for $\tau_X=10^8$~s. The dark (blue) shaded
regions correspond to an overabundance of D/H, i.e. regions where there
is net production of deuterium.  The lighter (blue) shaded regions
represent an underabundance of D/H or regions where there is net
destruction of deuterium.  Notice that the thin strips which are unshaded
for which the D/H abundance is acceptable.  These will be excluded
when the constraints from the other light elements are included.

\begin{figure}
\psfig{file=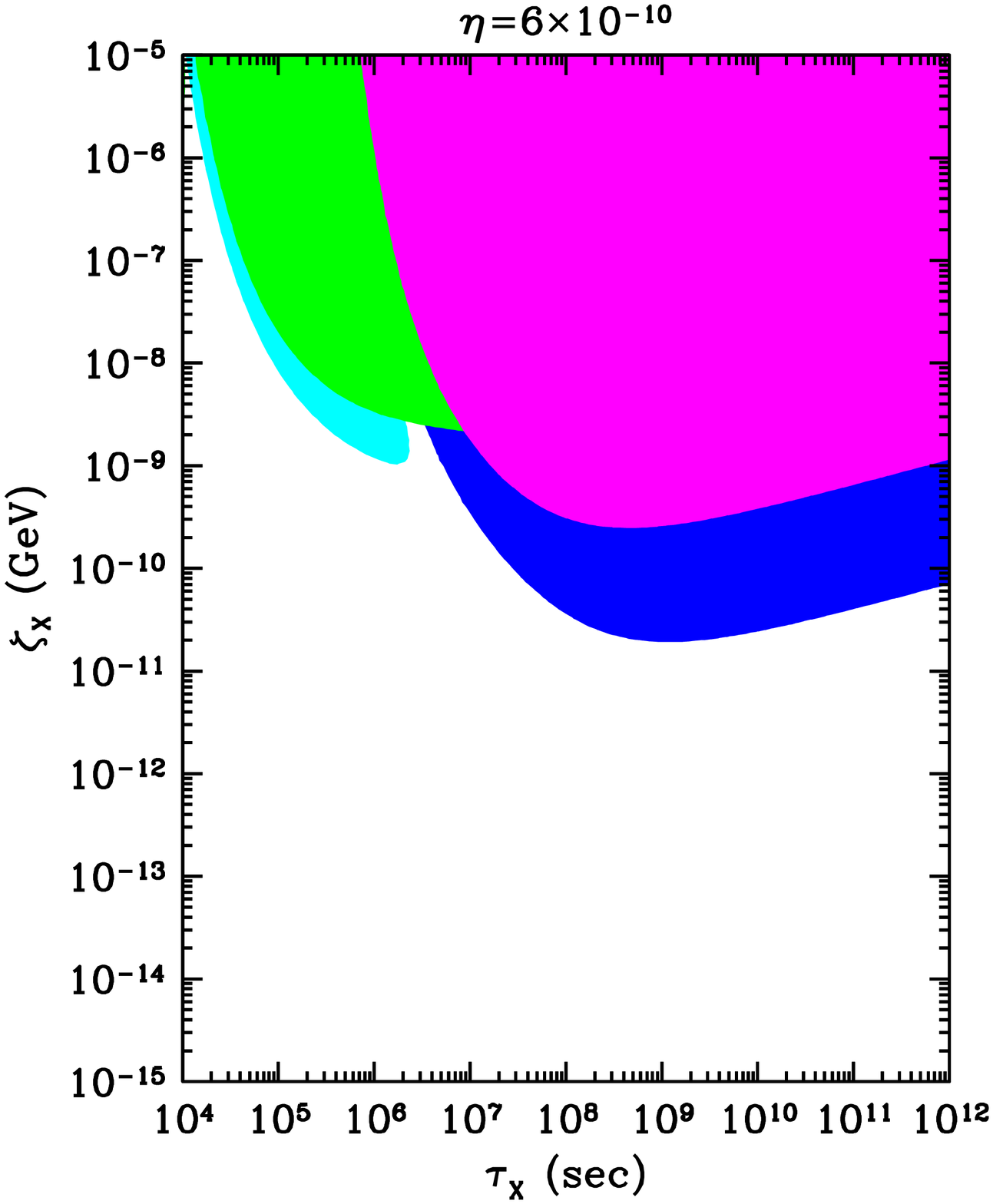,height=4in}
\hfill
\psfig{file=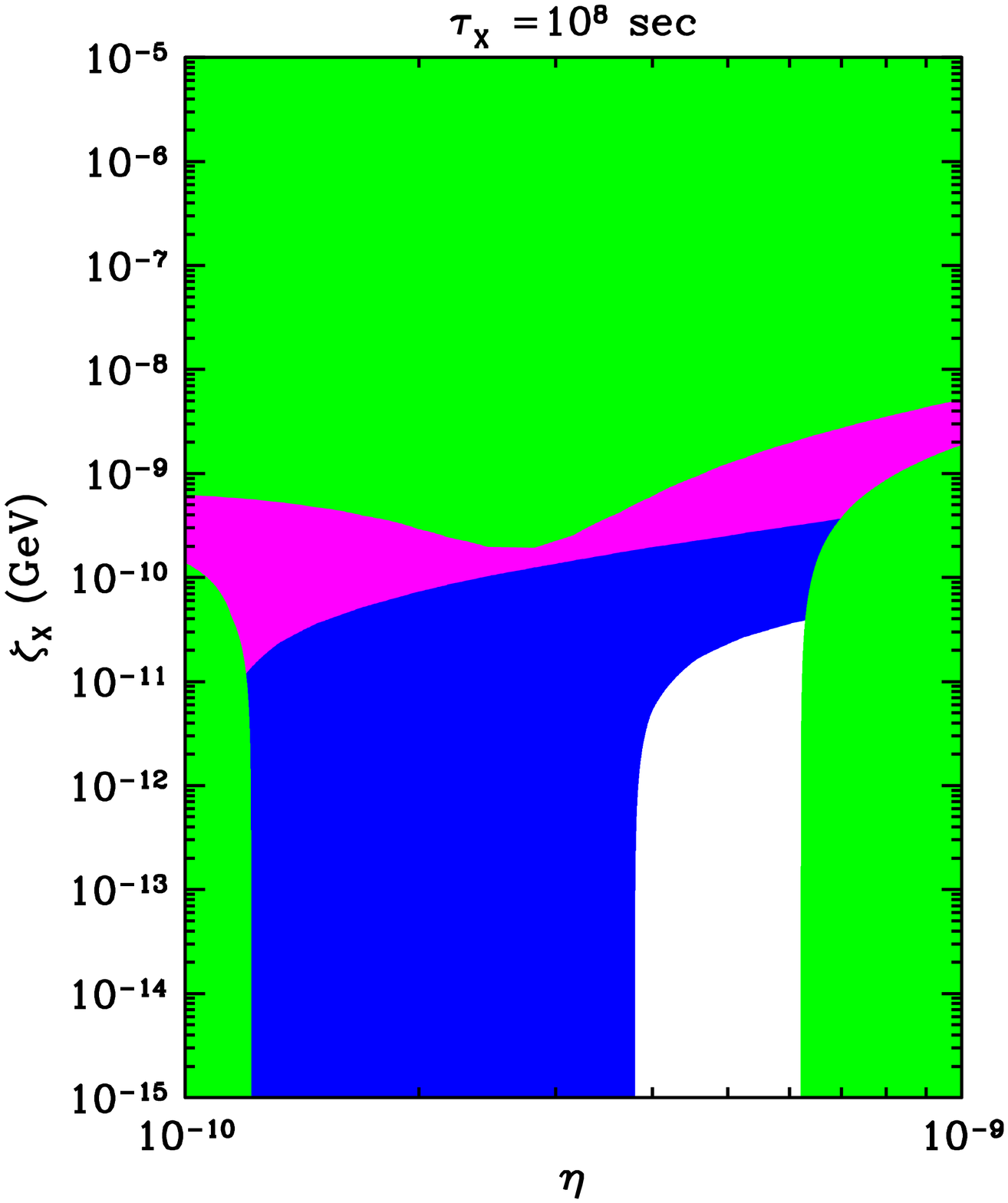,height=4in}
\caption{\it
As in Figure \protect\ref{fig:combined}, including the constraints from
\he4 - medium (pink) shading and \li7 - medium-light (green) shading.}
\label{fig:combined247}
\end{figure}

\begin{figure}
\psfig{file=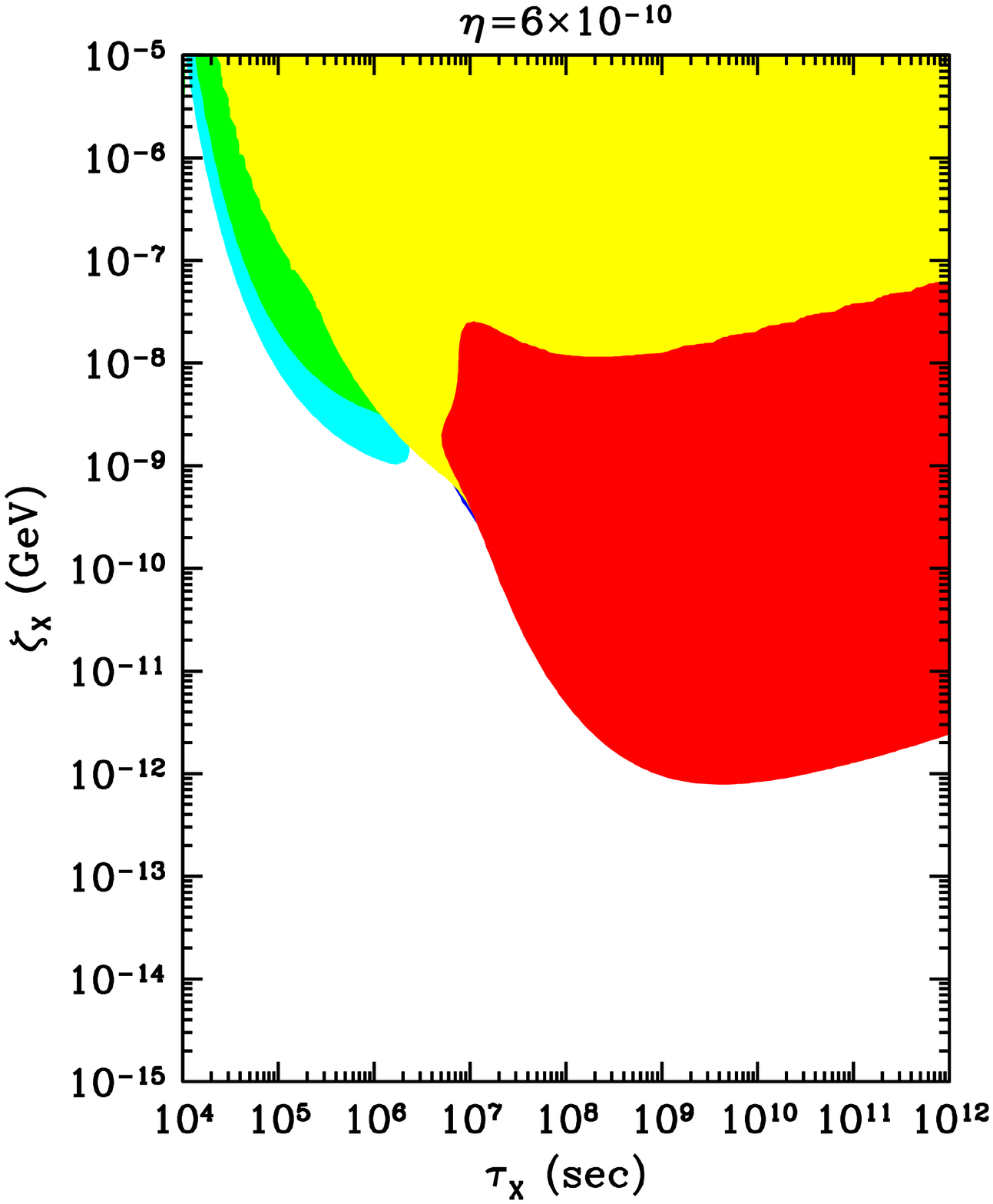,height=4in}
\hfill
\psfig{file=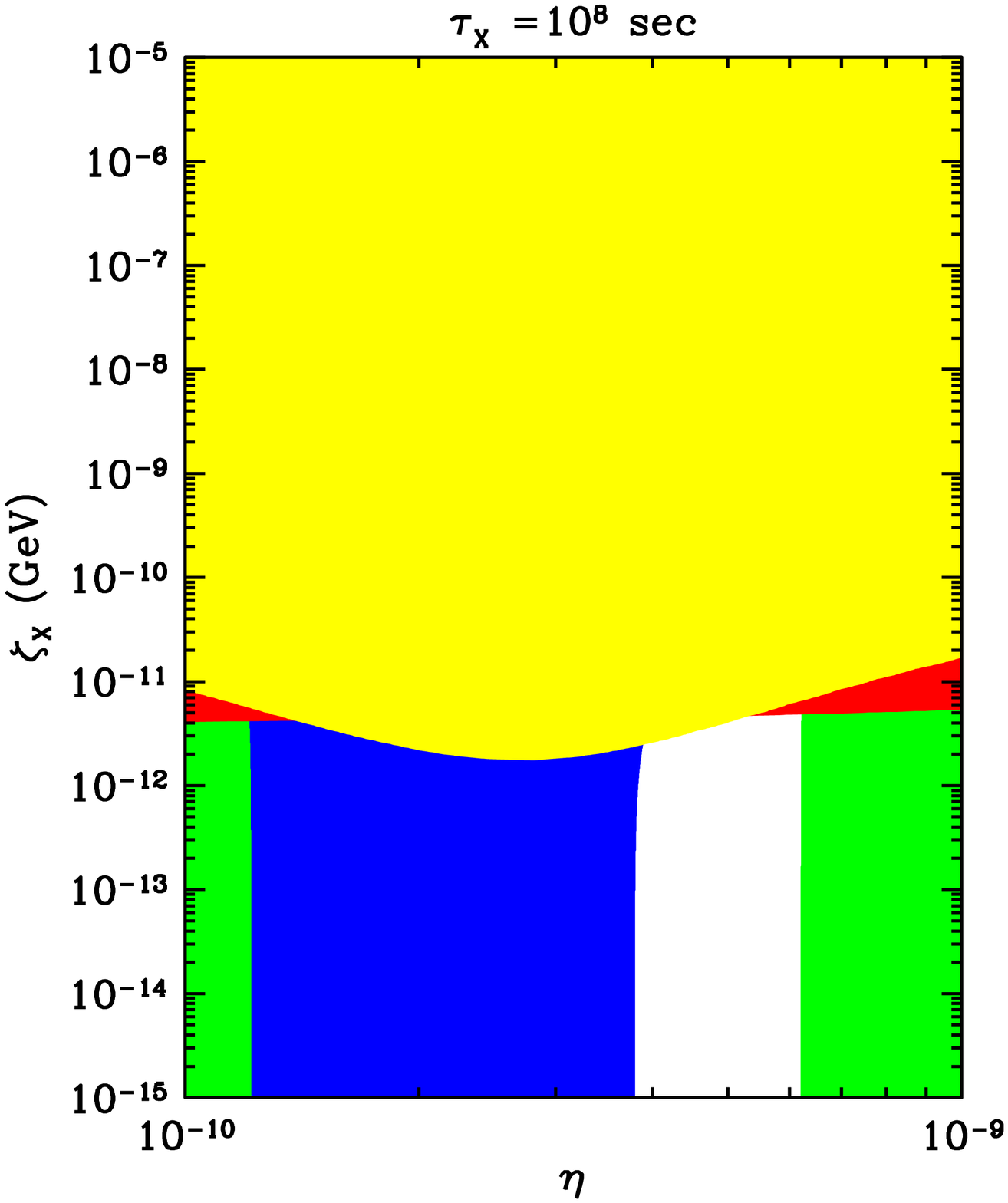,height=4in}
\caption{\it
As in Figures \protect\ref{fig:combined} and \protect\ref{fig:combined247}
including the constraints from
\li6 (dark red).}
\label{fig:combinedeta}
\end{figure}

In Figure \ref{fig:combined247}, we include the constraints from \he4 and
\li7.  Here we superimpose the \he4 constraint, shown as the medium
shaded (pink)
region. and the \li7 constraint, shown as the medium-light (green) shaded
region, on the D/H constraints. 
We see that D, \he4, and \li7
alone, i.e., primordial species, impose a limit of
\beq
\label{eq:Dlim}
\zeta_X \la \zeta_{\rm max} = \EE{3.5}{-11}\ {\rm GeV},
\eeq
which is dominated by the limits from D.
We can do better if we include \li6.
Our limit ${\rm \li6/H} \la 2 \times 10^{-12}$
pushes the above constraint down to 
\beq
\label{eq:Li6lim}
\zeta_X \la \zeta_{\rm max} = \EE{5}{-12} \ {\rm GeV}
\eeq
for $\tau_X = 10^{8}$ s as seen in Figure \ref{fig:combinedeta} by the
dark (red) shaded region.  The constraint from the \li6/\li7 ratio is
shown as the light (yellow) shaded region.  Notice that it becomes the
stronger constraint at $\eta < 5.0$.

These constraints are subject to uncertainties in the \li6 limit, due both
to the possible stellar depletion of \li6 and the known Galactic
production of \li6 by cosmic rays. Our limit is intended conservatively to
allow for both effects.  Even so, we see the power of \li6.  We thus urge
further observations of the Li isotopic ratio, as a firmer understanding
of this nuclide could further strengthen the constraint we have derived.

As already noted, if the observed light-element abundances retain their
current central values, but the error budget shrinks, then the
light-element data will be in discord with standard BBN.  Decaying
particles might provide one possible means of reconciling such
light-element observations and theory. As an illustrative example,
consider the case in which the CMB fixes $\eta_{10}=6$, and the observed
light element abundances remain as above, but with the total error budget
equal to that of the current statistical errors.  Then \li7 and D would be
in significant disagreement. One could, however, bring these nuclides and
\he4 into agreement by appealing to the decaying-particle scenario we have
laid out here, the allowed region of parameter space still open being one
in which a non-zero $\zeta_X$ is preferred. The new \li7 upper limit would
eliminate low values of $\zeta_X$, allowing only a narrow band with $\zeta
\approx10^{-9}$.  The observations would force us to live in the narrow
channel where D production and destruction are nearly balanced, with a
decaying particle lifetime $\tau_X \approx \EE{3}{6}$~sec.  In this
regime, \he4 is at its BBN value, because of its high photoerosion
threshold, shown in Table \ref{tab:photo-thresh}. However, the more weakly
bound \li7 is destroyed, at just the right level to bring the observations
in accord with the D observations.  This rather fine-tuned scenario is
testable with \li6 observations.  In this region of parameter space, we
predict a \li6 abundance of ${\rm \li6/H} \approx 10^{-11}$ and $\li6/\li7
\approx 0.03$. This \li6 abundance would appear as a pre-Galactic plateau
in halo stars.  Indeed, the \li6 level would be large enough to dominate
the cosmic-ray component of \li6 over most of the Population II
metallicity range.

\section{Application to Cosmological Gravitinos}
\label{sect:grav}

We now illustrate the impact of our calculations by discussing their
implications for cosmological gravitinos. In conventional supergravity 
scenarios, the gravitino is expected to have a mass comparable to that  
of supersymmetric partners of standard Model particles, which should 
weigh less than about 1~TeV if they are to stabilize the gauge hierarchy
\cite{heir}. 
Therefore, the gravitino is usually thought to weigh between about 
100~GeV to 10~TeV, though both larger and smaller masses have sometimes 
been considered. The lightest supersymmetric particle (LSP) is generally 
thought to be the lightest neutralino $\chi$, a model-dependent 
mixture of the photino ${\tilde \gamma}$, the zino ${\tilde Z}$ and 
the neutral Higgsinos ${\tilde H}_{1,2}$ \cite{EHNOS}. The LSP would be
stable in  models in which $R$ parity is conserved, as we assume here. On
the other  hand, the gravitino would be unstable, with a partial ${\tilde
G}
\to 
\chi + \gamma$ decay rate calculated to be
\begin{equation}
\Gamma (\tilde G \to \chi \gamma) \; = \; {1\over 4} \left( {M_{\tilde
G}^3
\over  M_{\rm P}^2} \right) O_{\chi \gamma}^2,
\label{gravrate}
\end{equation}
where $O_{\chi \gamma}$ is the fraction of ${\tilde \gamma}$ in the wave 
function of the LSP $\chi$, 
and $M_{\rm P}$ is the Planck mass. 
In many models, the LSP is essentially a pure 
$U(1)$ gaugino (Bino) ${\tilde B}$, in which case $O_{\chi \gamma} = 
\cos^2 \theta_W$ and the neutralino mixing does not suppress the gravitino 
decay rate (\ref{gravrate}). 
We assume for now that $\chi = {\tilde B}$ and that no other gravitino 
decay modes are significant, in which case the gravitino lifetime is
\begin{equation}
\tau_{\tilde G} \; = \; 2.9 \times 10^8 \left( {100~{\rm GeV} \over 
M_{\tilde G}} \right)^3~{\rm s}.
\label{gravlife}
\end{equation}
We discuss later the modifications to our analysis 
needed if the LSP is essentially a pure Higgsino, another possibility 
sometimes considered, or if other decay modes are open to the gravitino.

The production of gravitinos in the early Universe has been the subject of 
heated discussion. An unavoidable contribution is thermal production
\cite{EKN,Kawasaki:1994af,enor,buch}.  Here, we will apply our results in
combination with the recent calculation in~\cite{buch}, which gives
\begin{equation}
Y_{\tilde G} \equiv {n_{\tilde G} \over n_\gamma} \; = \; 1.2 \times
10^{-11} 
\left( 1 + {m_{\widetilde g}^2 \over 12 m_{\tilde G}^2}  \right) \times 
\left(
{T_R
\over 10^{10}~{\rm GeV}}
\right)
\label{gravY}
\end{equation}
where $T_R$ is the maximum temperature reached in the early
Universe\footnote{We note that this calculation is based on the dominant
strong contributions to gravitino production.  Electroweak corrections
would enhance the production rate by about 5 -- 20 \%.}. In  conventional
inflationary cosmology,
$T_R$ is the reheating temperature  achieved at the end of the
inflationary epoch. In some inflationary  scenarios, there may be
additional gravitino production, either during the  inflationary epoch or
later, before thermalization is achieved. Either of  these effects would
only accentuate the potential problem we discuss  below, and we do not
consider such possibilities here.

In (\ref{gravY}), $m_{\widetilde g}$ is the low-energy gluino mass.
In supersymmetric models with gaugino mass unification, there is a
definite
relation between the gluino mass and the bino mass, which, at the one-loop
order sufficient for our purposes, is $m_{\widetilde g}/m_{\widetilde B} 
\simeq
\alpha_3(m_{\widetilde g})/\alpha_1$.  Typically, this ratio is between 5
and 6. If the bino is the LSP, as we are presently considering,
then $m_{\tilde G} \ge m_{\widetilde B}$. The middle term in
(\ref{gravY}) is therefore never larger than 4, and tends to unity for 
large gravitino masses. Thus we estimate
\begin{equation}
Y_{\tilde G}  \; \simeq \; (0.7 - 2.7) \times
10^{-11} 
 \times \left(
{T_R
\over 10^{10}~{\rm GeV}}
\right)
\label{gravY2}
\end{equation}
in this case.

The upper limit (\ref{bestlimit}) can be expressed as a limit on
$Y_{\tilde G}$:
\begin{equation}
Y_{\tilde G}  < 5 \times
10^{-14} 
 \times \left(
{100~{\rm GeV}
\over m_{\tilde G}}
\right)
\label{limit}
\end{equation}
Comparing the calculated abundance (\ref{gravY2}) with the
upper limit (\ref{limit}), we infer the following upper limit on the
reheating temperature $T_R$, for $M_{\tilde G} \sim
100~{\rm  GeV}$:
\begin{equation}
T_R \; < \; (1.9 - 7.5) \times 10^7~{\rm GeV} 
\label{Tlimit}
\end{equation}
This upper limit is far smaller than the reheating temperature $T_R \sim 
10^{12}$~GeV expected in conventional inflationary scenarios. As noted in
the introduction, this bound places important constraints on models of
baryo/leptogenesis.

We recall that the upper limit (\ref{Tlimit}) comes from a combination of 
data
and calculations of different light-element abundances. It could not be
obtained by considering the deuterium abundance alone, as this would
allow a `tail' of the parameter space extending to large gravitino
abundances $Y_{\tilde G}$, as well as an isolated `channel' at large
$Y_{\tilde G}$. The `tail' cannot be excluded just by considering also the
abundance of \he4, though the `channel' probably can. However, as
discussed earlier, the `tail' can be excluded by also considering the
\li6 abundance. We have already discussed why we think that measurements
of the relevant photoreaction cross sections and the astrophysical data
are now sufficiently reliable for the \li6 data to be regarded as a
serious constraint.

As we commented at the end of the previous Section, if the observed
light-element abundances were to retain their current central values while
the error budget shrank, the light-element data would become inconsistent
with standard BBN, and decaying particles might be able to reconcile such
light-element observations with theory. This scenario would predict a
lifetime $\tau_{\tilde{G}} \approx \EE{3}{6}$~sec and an abundance
$\zeta_{\tilde{G}} \approx 10^{-9}$~GeV.  These constraints would place
the gravitino mass at $M_{\tilde{G}}\approx 460$~GeV, with a relic
abundance of $Y_{\tilde{G}}\approx \EE{2}{-12}$, corresponding in turn to
a reheating temperature $T_R \simeq \EE{(0.8 - 3.1)}{9}$~GeV.

{\it How might the potentially embarrassing conclusion (\ref{Tlimit}) be 
avoided or evaded?}

The first option one might consider is diluting the density of gravitinos
by several orders of magnitude some time between their production at a
temperature close to $T_R$ and the period when they decay. This large
entropy release should certainly occur before BBN,
i.e., when the age $t \lappeq 1$~s, in order to avoid destroying its
predictions completely. Very likely, such a large entropy release would
also have had to occur before baryogenesis, in order to avoid an
unacceptable dilution of the primordially-generated baryon asymmetry. The
latest epoch at which baryogenesis is seems likely to have occurred is the
electroweak phase transition, which occurred when the age $t \sim
10^{-10}$~s. Afleck-Dine baryogenesis \cite{ad} offers one such
possibility. In these models the Universe becomes dominated by the
oscillation of a scalar field along a supersymmetric flat direction. In
general, the net baryon asymmetry produced in these models can be quite
large, actually necessitating late entropy production \cite{aden}. The
dilution of the gravitino abundance would be an immediate consequence.  
The issue of gravitino production and dilution in connection with
baryogenesis was recently considered~\cite{Buonanno:2000cp} in the context
of the pre-big bang scenario~\cite{ven}.

It is possible that the mass and decay modes of the gravitino are such
that the bounds discussed here, which cover lifetimes from $10^4$ --
$10^{12}$ s, become inapplicable. 
We see from (\ref{gravrate}) that the rate for ${\tilde G} \to \chi +
\gamma$ decay would increase by three orders of magnitude if $M_{\tilde
G}$ were one order of magnitude larger. In fact, a heavier gravitino
might have additional decay modes open kinematically, possibly decreasing
its lifetime by another two orders of magnitude if all the MSSM particle
weighed less than $M_{\tilde G}$. If $\tau_{\tilde G} \sim 10^4$~s, as
might occur if
$M_{\tilde G} \sim 1$~TeV and it could decay into ${\tilde X_{EW}} +
X_{EW}$ as well as $\chi + \gamma$, where $X_{EW}$ is any Standard Model
particle with only electroweak interactions, our limits are significantly
weakened.
However, we remind the reader that
outside the range $10^4$ --
$10^{12}$ s other bounds come into play.  At smaller lifetimes,
hadronic decay products will upset the prediction of BBN
\cite{Reno:1987qw,Dimopoulos:1988ue,Kohri:2001jx}. {}From the recent
results of \cite{Kohri:2001jx}, one finds that for lifetimes in the range
1 -- $10^4$ s, the upper limit to $Y_{\tilde G}$ is of order $10^{-13}$,
and this bound weakens at lower lifetimes. For $\tau < {\rm few} \times
10^{-2}$ s, the BBN limit disappears. Similarly, at longer lifetimes $ >
10^{12}$ s, there are non-negligible constraints from the observed
gamma-ray background \cite{gamma,ellis,kr}. These are strongest for decay
lifetimes of order the age of the Universe $\sim 10^{18}$ s, where the
bound on $m_{\tilde G} Y_{\tilde G}$ is of order $10^{-16}$. At still
longer lifetimes, the bound weaks quickly and becomes ineffective at
lifetimes longer than about $10^{24}$ s.

Could the cosmological embarrassment be avoided if $M_{\tilde G} \ll
100$~GeV? In such a case, the gravitino would presumably be the LSP, and
absolutely stable if $R$ parity is conserved, as we have been assuming. In
fact, as argued in~\cite{buch}, a reheating temperature of order 
$10^{10}$~GeV would result in an acceptably large relic density of gravitinos.
Converting the gravitino abundance in (\ref{gravY2}) to its contribution
to closure density, one finds,
\begin{equation}
\Omega_{\tilde G} h^2 = (0.026 - 0.1)\times \left({m_{\tilde G} \over 100
\gev} \right) \left( {T_R \over 10^{10} \gev} \right)
\end{equation}
In this case, the lightest MSSM sparticle
(presumably the lightest neutralino
$\chi$) would be the next-to-lightest supersymmetric particle (NLSP), and
would itself be unstable: the decay NLSP $\to \gamma + {\tilde G}$ would
have a lifetime similar to (\ref{gravlife}), and the bound (\ref{gravY})
could be applied to the NLSP abundance $Y_{NLSP}$, which in terms of
$\Omega_\chi h^2$ is
\begin{equation}
\Omega_\chi h^2 < 2 \times 10^{-4}.
\label{nolsp}
\end{equation} 
However, suppressing
$\Omega_\chi h^2$ to such a low value seems very difficult. Characteristic
values of $\Omega_\chi h^2$ in the MSSM are ${\cal O}(0.01 - 10.0)$
\cite{eos}, corresponding to $Y_\chi \sim 10^{-10}$, which is far above 
the bound
(\ref{gravY}). Indeed, in much of the phenomenologically allowed 
supersymmetric
parameter space, the relic density is too large. Relic neutralino
densities as low as (\ref{nolsp}) would require special parameter choices
for which the neutralino is either primarily a massive Higgsino,
in which case important coannihilation effects might suppress the 
abundance, or the neutralino mass happens to be very close to half the 
mass of the Heavy Higgs scalar
and pseudo scalar, so that rapid $s$-channel annihilation reduces the 
neutralino density. In the former case, it is in fact unlikely that 
coannihilations are sufficiently strong to satisfy the bound
(\ref{nolsp}), despite reducing the relic density to a level well below
critical. In the latter case, however, very low relic densities are 
possible.

Could the cosmological embarrassment be avoided by altering the
composition of the LSP into which the gravitino is supposed to decay? The
gravitino decay amplitude could be suppressed, and the lifetime estimate
(\ref{gravlife}) correspondingly increased, if the LSP did not have
photino component, for example if it were essentially a pure Higgsino.  
However, in view of the discussion above, in realistic models it seems
difficult to suppress the
${\tilde G} - \chi - \gamma$ coupling enough to increase the gravitino
lifetime sufficiently to relax the bound (\ref{gravY})
adequately, bearing in mind the bounds applicable for longer 
lifetimes.

Finally, we consider briefly the situation if $R$ parity is violated.  
First we consider $R$-violating decays of the lightest MSSM particle,
assumed to be the lightest neutralino $\chi$. As discussed earlier, its
abundance is likely to be $Y_\chi \gappeq 10^{-10}$, which conflicts with
our bounds unless $\tau_\chi \lappeq 10^4$~s~\footnote{Note that the case
$\tau_\chi \lappeq 10^2$~s requires further consideration, going beyond 
the
scope of this paper.}. Considering now the decays of the gravitino,
assuming it to be heavier than $\chi$, there are two options to consider.  
The simplest possibility is that it decays in the same way as in the
$R$-conserving case: ${\tilde G} \to \chi + \gamma$, etc., in which case
the previous $R$-conserving gravitino analysis applies, and we again
conclude that $\tau_{\tilde G} \lappeq 10^4$~s. Alternatively, the 
dominant
${\tilde G}$ decays might violate $R$ parity. In this case, the bound
$\tau_\chi \lappeq 10^4$~s again applies, and, as discussed above, the 
same
bound applies to the decays of the lightest MSSM sparticle.

\section{Discussion and Conclusions}

We have re-examined in this paper the upper limits on the possible
abundance of any unstable massive relic particle that are provided by the
success of Big-Bang Nucleosynthesis calculations. A new aspect of this
work has been the use of cosmic microwave background data to constrain
independently the baryon-to-photon ratio, which was not possible in
previous studies of this problem. We have also incorporated in our
analysis the an updated suite of photonuclear and nuclear
cross sections, and calculated both
analytically and numerically the network of reactions induced by
electromagnetic showers that create and destroy the light elements
deuterium, \he3, \he4, \li6 and \li7.

It was pointed out in previous work that considerations of the deuterium
abundance alone would allow certain exceptional regions of parameter space
with relatively large abundances of unstable particles. However, as shown
in this paper, considerations of the abundances of \he4 and \li6 exclude
these particular regions.

We have illustrated our results by applying them to massive gravitinos. If
they weigh $\sim 100$~GeV, their primordial abundance $Y_{\tilde G}$
should have been $\lappeq 5 \times 10^{-14} \times \left({100~{\rm GeV} /
m_{\tilde G}}\right)$, corresponding to a reheating temperature $T_R \;  
< \; (1.9 - 7.5) \times 10^7$~GeV. This could present a potential
difficulty for some models of inflation and leptogenesis. We have
discussed various scenarios for evading this potential embarrassment, for
example by varying the gravitino mass, or by postulating an alternative
scenario for baryogenesis, such as non-thermal leptogenesis or the
Affleck-Dine mechanism.

This example of the gravitino illustrates the power and importance of the
cosmological upper limits on the abundances of unstable massive particles.  
Extensions of this analysis are clearly desirable. For example, it would
be valuable to combine our analysis of electromagnetic decay cascades with
a similar analysis of hadronic showers, a topic that lies beyond the scope
of this paper.

\subsection*{Acknowledgments}

\noindent 
We would like to thank Wilfried Buchm\"uller for helpful communications. 
The work of K.A.O.\ was partially
supported by DOE grant DE--FG02--94ER--40823 at the Universoty of
Minnesota. This work of BDF and RHC was supported 
by NSF grant AST 00-92939 at the University of Illinois.

\appendix

\section{Cross Sections}
\label{sect:cs}

Kawasaki, Kohri and Moroi~\cite{kkm} have provided a useful table of
reactions and references to nuclear data.  We have supplemented the data
by using tabulations made by the National Nuclear Data Center
(NNDC)~\cite{nndc} and the NACRE collaboration~\cite{nacre}. The relevant
cross sections are listed here for convenience, and we refer the
interested reader to the NNDC and NACRE websites for further references on
the data.

We have computed thresholds and $|Q|$ values using the mass data of Audi
and Wapstra\cite{audi}, available on the US Nuclear Data Program
website~\cite{USNDP}. Reverse reaction
data were sometimes available, in which cases we used detailed balance to 
transform
the data into forward data.  The equations of detailed balance for the
reactions $\gamma + T \rightarrow A + B$ and $P+T\rightarrow A + B$ are:
\beqar
\sigma_{\gamma + T \rightarrow A + B} &=&
\frac{g_Ag_B}{(1+\delta_{AB})g_T} \left(
\frac{\mu E_{cm}(A,B)}{E_\gamma^2} \right) \sigma_{A+B\rightarrow T+\gamma} \\
\sigma_{P+T\rightarrow A+B} &=&
\frac{(1+\delta_{PT})g_Ag_Bm_Am_BE_{cm}(A,B)}{
(1+\delta_{AB})g_Pg_Tm_Pm_TE_{cm}(P,T)} \sigma_{A+B\rightarrow P+T},
\eeqar
where the $g_i$ are the statistical weights of each species, $\mu $ is the
reduced mass of the system $A+B$, and $E_{cm}(x,y)$ is the center-of-mass
energy of the system $x+y$.  See Blatt and Weisskopf~\cite{blatt} and
Fowler, Caughlan, and Zimmerman~\cite{fcz} for discussions on these
relations.

In the numerical fits, all energies are in MeV.
In a few cases, which we have noted, the fits are
those previously published.
Otherwise, we have adopted a specific empirical form for the
nonresonant parts of the cross sections.
This form is the product of a power law in 
photon energy $E_\gamma$, and a power law in 
photon energy above threshold, $E_\gamma - |Q|$.
We have found that expressions of this type provide
a simple but accurate representation of the data.

\begin{enumerate}

\item
d($\gamma$,n)p \ \ \ \ \ $E_{\gamma,th} = |Q| = 2.224573$ MeV~\cite{dgnp}.
\bdm
\!\!\!\!\!\!\sigma (E_\gamma) = 18.75 {\rm mb}\!\!\left[ \left(
\frac{\sqrt{|Q|(E_\gamma - |Q|)}}{E} \right)^{\!\!3}\!\! + 0.007947
\left( \frac{\sqrt{|Q|(E_\gamma - |Q|)}}{E} \right)^{\!\!2}\!\! \frac{\left(
\sqrt{|Q|} - \sqrt{0.037} \right)^2}{E_\gamma - (|Q| - 0.037)} \right]
\edm

\item
t($\gamma$,n)d \ \ \ \ \ $E_{\gamma,th} = |Q| = 6.257248$ 
MeV~\cite{tg1,tg2}.
\bdm
\!\!\!\!\!\!\sigma (E_\gamma) = 9.8 {\rm mb} \frac{|Q|^{1.95}(E_\gamma
-|Q|)^{1.65}}{E_\gamma^{3.6}} 
\edm

\item
t($\gamma$,np)n \ \ \ \ \ $E_{\gamma,th} = |Q| = 8.481821$ MeV~\cite{tg2}.
\bdm
\!\!\!\!\!\!\sigma (E_\gamma) = 26.0 {\rm mb} \frac{|Q|^{2.6}(E_\gamma -
|Q|)^{2.3}}{E_\gamma^{4.9}}
\edm

\item
\he3($\gamma$,p)d \ \ \ \ \ $E_{\gamma,th} = |Q| = 5.493485$ 
MeV~\cite{3g1,3g2}. We use reverse reaction data from
NACRE~\cite{nacre,r3g1}.
\bdm
\!\!\!\!\!\!\sigma (E_\gamma) = 8.88 {\rm mb} \frac{|Q|^{1.75}(E_\gamma
-|Q|)^{1.65}}{E_\gamma^{3.4}} 
\edm

\item
\he3($\gamma$,np)p \ \ \ \ \ $E_{\gamma,th} = |Q| = 7.718058$ 
MeV~\cite{3g2,3g2pn1}.

\bdm
\!\!\!\!\!\!\sigma (E_\gamma) = 16.7 {\rm mb} \frac{|Q|^{1.95}(E_\gamma
- |Q|)^{2.3}}{E_\gamma^{4.25}} 
\edm

\item
\he4($\gamma$,p)t \ \ \ \ \ $E_{\gamma,th} = |Q| = 19.813852$ 
MeV~\cite{4gpt1,4gpt2,4gpt3}.

\bdm
\!\!\!\!\!\!\sigma (E_\gamma) = 19.5 {\rm mb} \frac{|Q|^{3.5}(E_\gamma -
|Q|)^{1.0}}{E_\gamma^{4.5}} 
\edm

\item
\he4($\gamma$,n)\he3 \ \ \ \ \ $E_{\gamma,th} = |Q| = 20.577615$ 
MeV~\cite{4gpt2,4gn31,4gn32,4gn33,4gn34}.

\bdm
\!\!\!\!\!\!\sigma (E_\gamma) = 17.1 {\rm mb} \frac{|Q|^{3.5}(E_\gamma -
|Q|)^{1.0}}{E_\gamma^{4.5}}
\edm

\item
\he4($\gamma$,d)d \ \ \ \ \ $E_{\gamma,th} = |Q| = 23.846527$ 
MeV~\cite{4gdd1,4gdd2}. We use reverse reaction data from
NACRE~\cite{nacre,r4gdd}.

\bdm
\!\!\!\!\!\!\sigma (E_\gamma) = 10.7 {\rm mb} \frac{|Q|^{10.2}(E_\gamma
- |Q|)^{3.4}}{E_\gamma^{13.6}} 
\edm

\item
\he4($\gamma$,np)d \ \ \ \ \ $E_{\gamma,th} = |Q| = 26.0711$ 
MeV~\cite{4gn31}.
\bdm
\!\!\!\!\!\!\sigma (E_\gamma) = 21.7 {\rm mb} \frac{|Q|^{4.0}(E_\gamma -
|Q|)^{3.0}}{E_\gamma^{7.0}}
\edm

\item
\li6($\gamma$,np)\he4 \ \ \ \ \ $E_{\gamma,th} = |Q| = 3.698892$ 
MeV~\cite{6g41,6g42}.

\bdm
\!\!\!\!\!\!\sigma (E_\gamma) = 104. {\rm mb} \frac{|Q|^{2.3}(E_\gamma -
|Q|)^{4.7}}{E_\gamma^{7.0}}
\edm

\item
\li6($\gamma$,X)$^3$A \ \ \ \ \ $E_{\gamma,th} = |Q| = 15.794685$ 
MeV~\cite{6g3X}.

\bdm
\!\!\!\!\!\!\sigma (E_\gamma) = 38.1 {\rm mb} \frac{|Q|^{3.0}(E_\gamma -
|Q|)^{2.0}}{E_\gamma^{5.0}}\times
\edm
\bdm
\!\!\!\!\!\!\left( 3.7\exp{\left[-\frac{1}{2}\left(
\frac{E_\gamma - 19.0}{3.5} \right)^2\right]} +
2.75\exp{\left[-\frac{1}{2}\left( 
\frac{E_\gamma - 30.0}{3.0} \right)^2\right]} +
2.2\exp{\left[-\frac{1}{2}\left( 
\frac{E_\gamma - 43.0}{5.0} \right)^2\right]} \right)
\edm

\item
\li7($\gamma$,t)\he4 \ \ \ \ \ $E_{\gamma,th} = |Q| = 2.467032$ MeV \
\ \ \ \ $E_{cm} = E_\gamma - |Q|$~\cite{6g42}. We use reverse reaction 
data from NACRE~\cite{nacre}, with 
modifications from Cyburt, Fields, and Olive~\cite{cfo1,rli7ga}.

\bdm
\!\!\!\!\!\!\sigma (E_\gamma) = 0.105 {\rm mb}\left(
\frac{2371 {\rm MeV^2}}{E_\gamma^2} \right) \exp{\left(
\frac{-2.5954}{\sqrt{E_{cm}}}\right)}\times
\edm
\bdm
\exp{(-2.056E_{cm})} 
\left( 1. + 2.2875E_{cm}^2 - 1.1798E_{cm}^3 + 2.5279E_{cm}^4\right)
\edm

\item
\li7($\gamma$,n)\li6 \ \ \ \ \ $E_{\gamma,th} = |Q| = 7.249962$ 
MeV~\cite{li7g61,li7g62,li7g63}.

\bdm
\!\!\!\!\!\!\sigma (E_\gamma) = 0.176 {\rm mb} \frac{|Q|^{1.51}(E_\gamma -
|Q|)^{0.49}}{E_\gamma^{2.0}} + 1205 {\rm mb} \frac{|Q|^{5.5}(E_\gamma -
|Q|)^{5.0}}{E_\gamma^{10.5}} + \frac{0.06 {\rm mb}}{1 + \left(
\frac{E_{cm} - 7.46}{0.188} \right)^2}
\edm

\item
\li7($\gamma$,2np)\he4 \ \ \ \ \ $E_{\gamma,th} = |Q| = 10.948850$ 
MeV~\cite{6g42}.

\bdm
\!\!\!\!\!\!\sigma (E_\gamma) = 122. {\rm mb} \frac{|Q|^{4.0}(E_\gamma -
|Q|)^{3.0}}{E_\gamma^{7.0}}
\edm

\item
\be7($\gamma$,\he3)\he4 \ \ \ \ \ $E_{\gamma,th} = |Q| = 1.586627$ MeV
\ \ \ \ \ $E_{cm} = E_\gamma - |Q|$. We use reverse reaction data from 
NACRE~\cite{nacre}, with
modifications from Cyburt, Fields, and Olive~\cite{cfo1,rbe7ga}.

\bdm
\!\!\!\!\!\!\sigma (E_\gamma) = 0.504 {\rm mb}\left(
\frac{2371 {\rm MeV^2}}{E_\gamma^2} \right) \exp{\left(
\frac{-5.1909}{\sqrt{E_{cm}}}\right)}\times
\edm
\bdm
\exp{(-0.548E_{cm})} 
\left( 1. - 0.428E_{cm}^2 + 0.534E_{cm}^3 - 0.115E_{cm}^4\right)
\edm

\item
\be7($\gamma$,p)\li6 \ \ \ \ \ $E_{\gamma,th} = |Q| = 5.605794$ MeV. 
We use reverse reaction data from NACRE~\cite{nacre,rbe7g6}.

\bdm
\!\!\!\!\!\!\sigma (E_\gamma) = 32.6 {\rm mb} \frac{|Q|^{10.0}(E_\gamma -
|Q|)^{2.0}}{E_\gamma^{12.0}} + \EE{2.27}{6} {\rm mb}
\frac{|Q|^{8.8335}(E_\gamma - |Q|)^{13.0}}{E_\gamma^{21.8335}}
\edm

\item
\be7($\gamma$,2pn)\he4 \ \ \ \ \ $E_{\gamma,th} = |Q| = 9.30468$ MeV. No 
data exist.  We assume isospin symmetry and use the
data for the reaction $\li7(\gamma, 2np)\he4$~\cite{6g42}.

\bdm
\!\!\!\!\!\!\sigma (E_\gamma) = 133. {\rm mb} \frac{|Q|^{4.0}(E_\gamma -
|Q|)^{3.0}}{E_\gamma^{7.0}}
\edm


\item
\he4(t,n)\li6 \ \ \ \ \ $E_{p,th} = 8.386972$ MeV \ \ $|Q| =
4.78293$ MeV \ \ $E_{cm} = \frac{m_4}{m_4 + m_t} E_p$. We use reverse 
reaction data~\cite{li6nat}.

\bdm
\!\!\!\!\!\!\sigma (E_\gamma) = 1940. {\rm mb} \frac{[(E_{cm} -
|Q|)/|Q|]^{.75}}{1 + \left( \frac{E_{cm} - 5.03043}{.09} \right)^2} +
46.2 {\rm mb} \frac{[(E_{cm} -
|Q|)/|Q|]^{.25}}{1 + \left( \frac{E_{cm} - 7.1329}{3.0} \right)^2} \\
+  32.8 {\rm mb}  \frac{[(E_{cm} -
|Q|)/|Q|]^{.5}}{1 + \left( \frac{E_{cm} - 7.5239}{25.0} \right)^2} 
\edm

\begin{figure}
\begin{center}
\epsfig{file=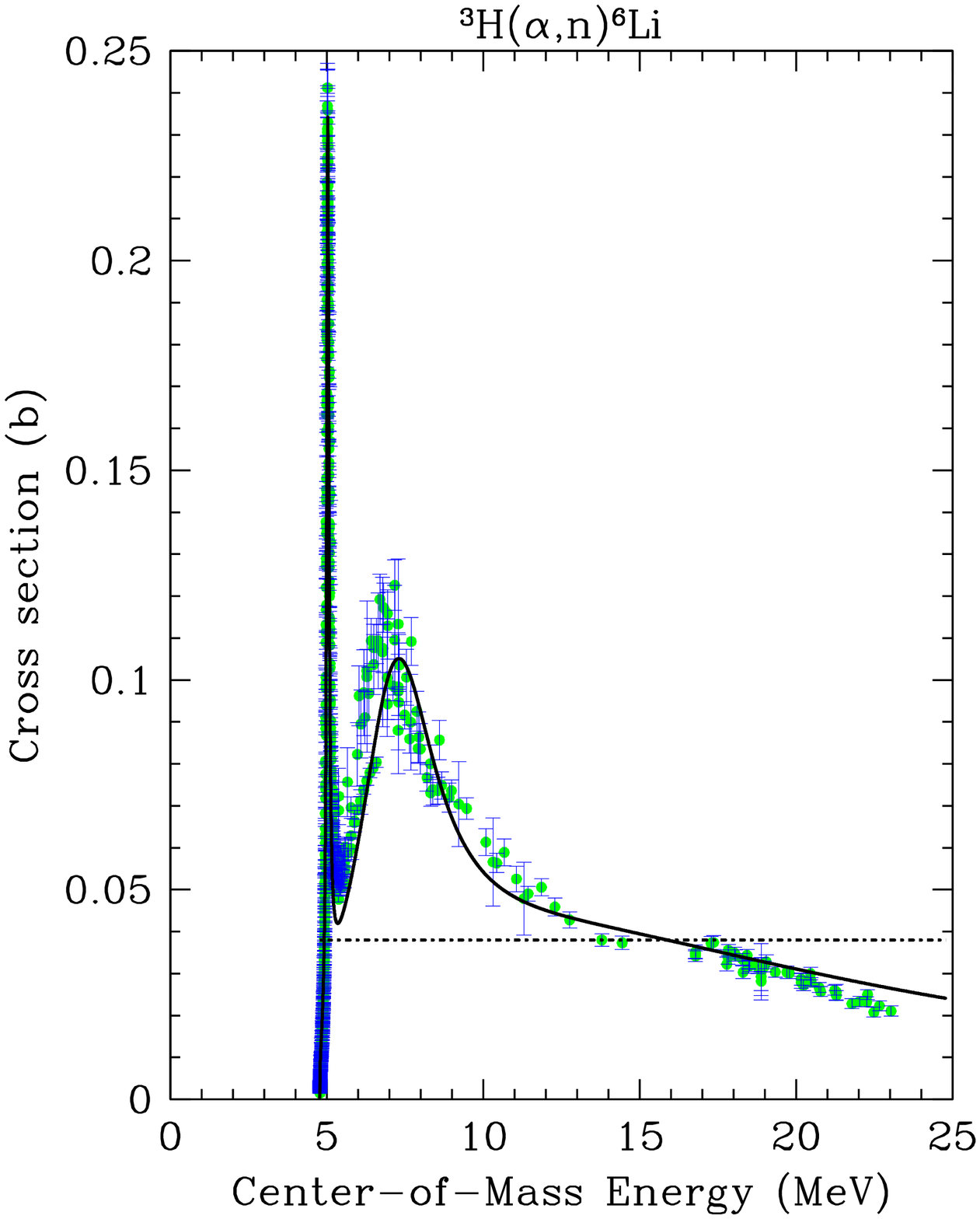,height=4.5in}
\end{center}
\vskip -.3in
\caption{\it
Cross-section data for the reaction  \he4(t,n)\li6 are plotted versus the
center-of-mass energy.  The solid curve is the 
parameterization we use, whilst the dashed curve is that adopted 
in previous studies~\cite{kartsen,kkm}.}
\label{fig:he4tn6}
\end{figure}

\item
\he4(\he3,p)\li6 \ \ \ \ \ $E_{p,th} = 7.047667$ MeV \ \ $|Q| =
4.019167$ MeV \ \ $E_{cm} = \frac{m_4}{m_4 + m_3} E_p$. We use reverse 
reaction data~\cite{li6pa3}.

\bdm
\!\!\!\!\!\!\sigma (E_\gamma) = 170. {\rm mb} \frac{[(E_{cm} -
|Q|)/|Q|]^{.75}}{1 + \left( \frac{E_{cm} - 5.8192}{.6} \right)^2} +
75.0 {\rm mb} \frac{[(E_{cm} -
|Q|)/|Q|]^{.25}}{1 + \left( \frac{E_{cm} - 7.5192}{3.0} \right)^2} \\
+ 32.1 {\rm mb}  \frac{[(E_{cm} -
|Q|)/|Q|]^{.5}}{1 + \left( \frac{E_{cm} - 12.019}{11.5} \right)^2} 
\edm

\begin{figure}
\begin{center}
\epsfig{file=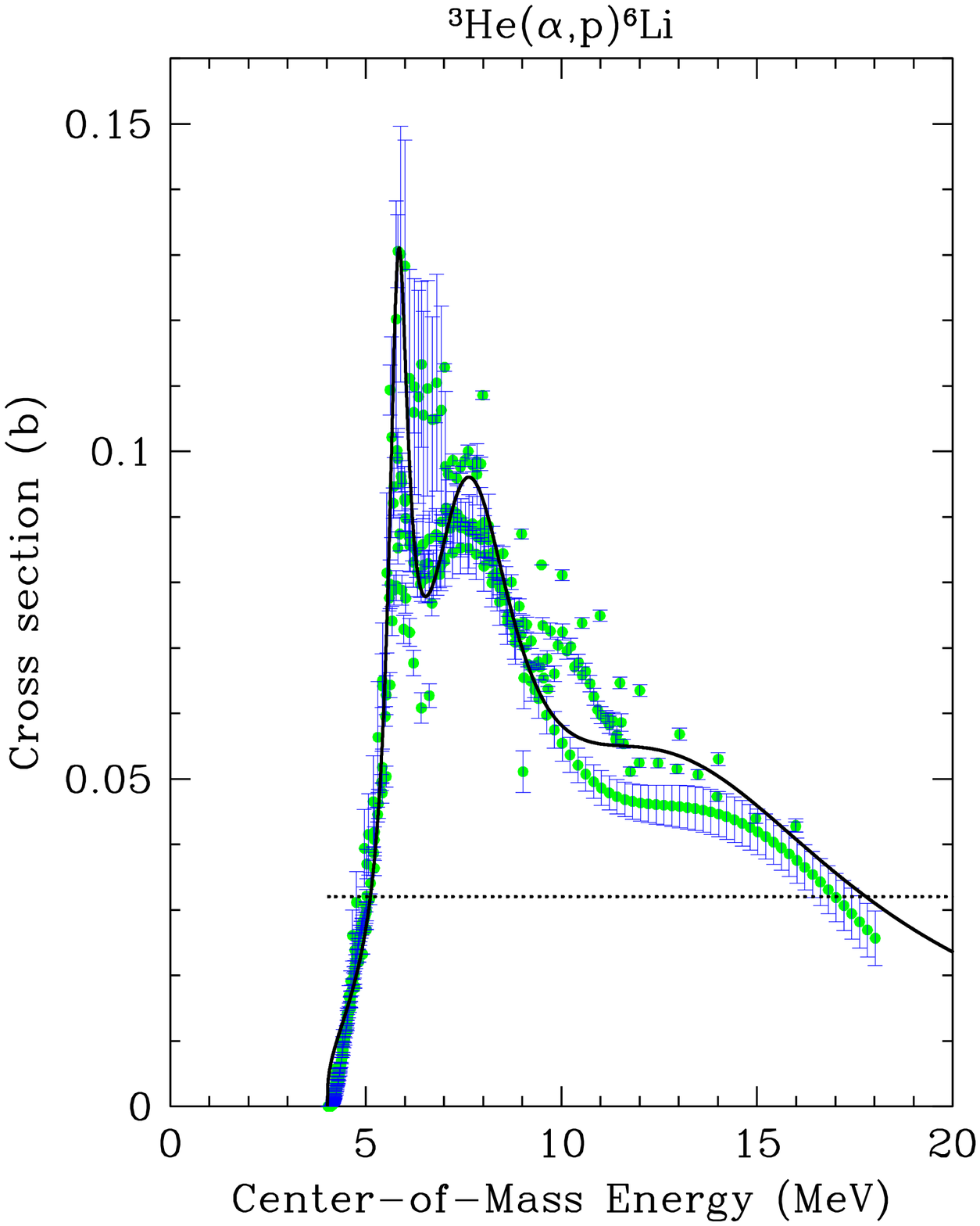,height=4.5in}
\end{center}
\vskip -.3in
\caption{\it
{Cross-section data for the reaction \he4(\he3,p)\li6 are plotted versus 
the center-of-mass energy. The solid curve is the parametrization we use, 
whilst the dashed curve is that adopted in previous 
studies~\cite{kartsen,kkm}.}
}
\label{fig:he4he3p6}
\end{figure}

\end{enumerate}


\end{document}